\documentclass[%
 preprint,
 amsmath,amssymb,
prstab,
]{revtex4-1}

\usepackage{graphicx}
\usepackage{dcolumn}
\usepackage{bm}
\usepackage{soul}
\usepackage[dvipsnames]{xcolor}

\begin{document}

\title{Convergence map with action-angle variables based on square matrix
for nonlinear lattice optimization}
\author{Li Hua Yu}
\author{Yoshiteru Hidaka}
\author{Victor Smaluk}
\affiliation{Brookhaven National Laboratory, Upton, New York 11973, USA}
\author{Kelly Anderson}
\author{Yue Hao}
\affiliation{Michigan State University, East Lansing, Michigan 48824, USA
}%
\begin{abstract}
To analyze nonlinear dynamic systems, we developed a new technique
based on the square matrix method. We propose this technique called
the ``convergence map'' for generating particle
stability diagrams similar to the frequency maps widely used in accelerator
physics to estimate dynamic aperture. The convergence map provides
similar information as the frequency map but in a much shorter computing
time. The dynamic equation can be rewritten in terms of action-angle
variables provided by the square matrix derived from the accelerator
lattice. The convergence map is obtained by solving the exact nonlinear
equation iteratively by the perturbation method using Fourier transform
and studying convergence. When the iteration is convergent, the solution
is expressed as a quasi-periodic analytical function as a highly accurate
approximation, and hence the motion is stable. The border of stable
motion determines the dynamical aperture. As an example, we applied
the new method to the nonlinear optimization of the NSLS-II storage
ring and demonstrated a dynamic aperture comparable to or larger than
the nominal one obtained by particle tracking. The computation speed
of the convergence map is 30 to 300 times faster than the speed of
the particle tracking, depending on the size of the ring lattice (number
of superperiods). The computation speed ratio is larger for complex
lattices with low symmetry, such as particle colliders. 
\end{abstract}
\maketitle

\section{Introduction}

The field of nonlinear dynamics has a very wide area of application
in science \cite{lieberman}. One of the topical applications
is to study the question of the long-term behavior of charged particles
in storage rings. One would like to analyze particle behavior under
many iterations of the one-turn map. The most accurate and reliable
numerical approach is particle tracking in a magnet lattice model
with appropriate integration methods. This approach is implemented
in many computer codes. However, particle tracking is computing resource-intensive,
so parallel codes and long computation time are often required. For
fast analysis, however, one would like a more compact representation
of the one-turn map out of which to extract relevant information.
Among the many approaches to this issue, we may mention canonical
perturbation theory, Lie operators, power series, normal form \cite{lieberman,ruth,guignard,schoch,dragt,berz,chao,bazzani,forest,forest1,michelotti} 
, etc. The results are often expressed as polynomials. However, for
increased perturbation, near resonance, or for large oscillation amplitudes,
these perturbative approaches often have insufficient precision. The
stability analysis of the beam trajectory and calculation of the dynamic
aperture requires an accurate solution of the nonlinear dynamical
equation. Hence there is a need to extract the information about long-term
particle behavior from the one-turn map based on these polynomials
with high precision and high speed.

The square matrix analysis \cite{yu1,yu2,yu3,hao} has a good
potential to explore this area. In this paper, we introduce a ``convergence
map'' calculated using action-angle variables in
the form of polynomials provided by a square matrix, which is derived
from the one-turn map for an accelerator lattice. Since the iterations
leading to the solution of the nonlinear dynamic equations expressed
by these action-angle variables can be carried
out by Fourier transform, the computation speed is very high, the
details are presented in Section II. Using the NSLS-II lattice \cite{NSLSII}
as an example, we show the nonlinear lattice optimization using the
convergence map results in a dynamic aperture comparable to or larger
than that obtained by particle tracking but the calculations are much
faster. In comparison with the frequency map \cite{laskar}
calculated by particle tracking, the convergence map is different,
even though it provides nearly the same information about the stable
region of the motion but the computation time is shorter by a factor
of 30 to 300 depending on the size and order of symmetry of the ring
lattice (number of superperiods). The computation speed ratio is larger
for complex lattices with low symmetry, such as particle colliders. 

As an example, Figure 1 shows a comparison of the convergence map
(a) and the frequency map from tracking (b) calculated with the same
number of points in the horizontal (x) and vertical (y) plane for
the nominal lattice of NSLS-II (1 superperiod). The computation speed
ratio is about 30 for this case. Figure 2 represents the computation
time of the convergence map and the frequency map as a function of
the number of points in both planes for one superperiod and the whole
NSLS-II ring consisting of 15 superperiods. 

One point we found is that the convergence map is more time-efficient
because the lattice model is represented by truncated power series
(TPS) \cite{berz, PyTPSA} and this time-consuming calculation is done only
once before the map generation, unlike the frequency map which requires
the tracking through the full lattice for every point on the map.The
details will be explained in the following sections. 

\begin{figure}[!htbp]
\includegraphics[width=0.45\textwidth]{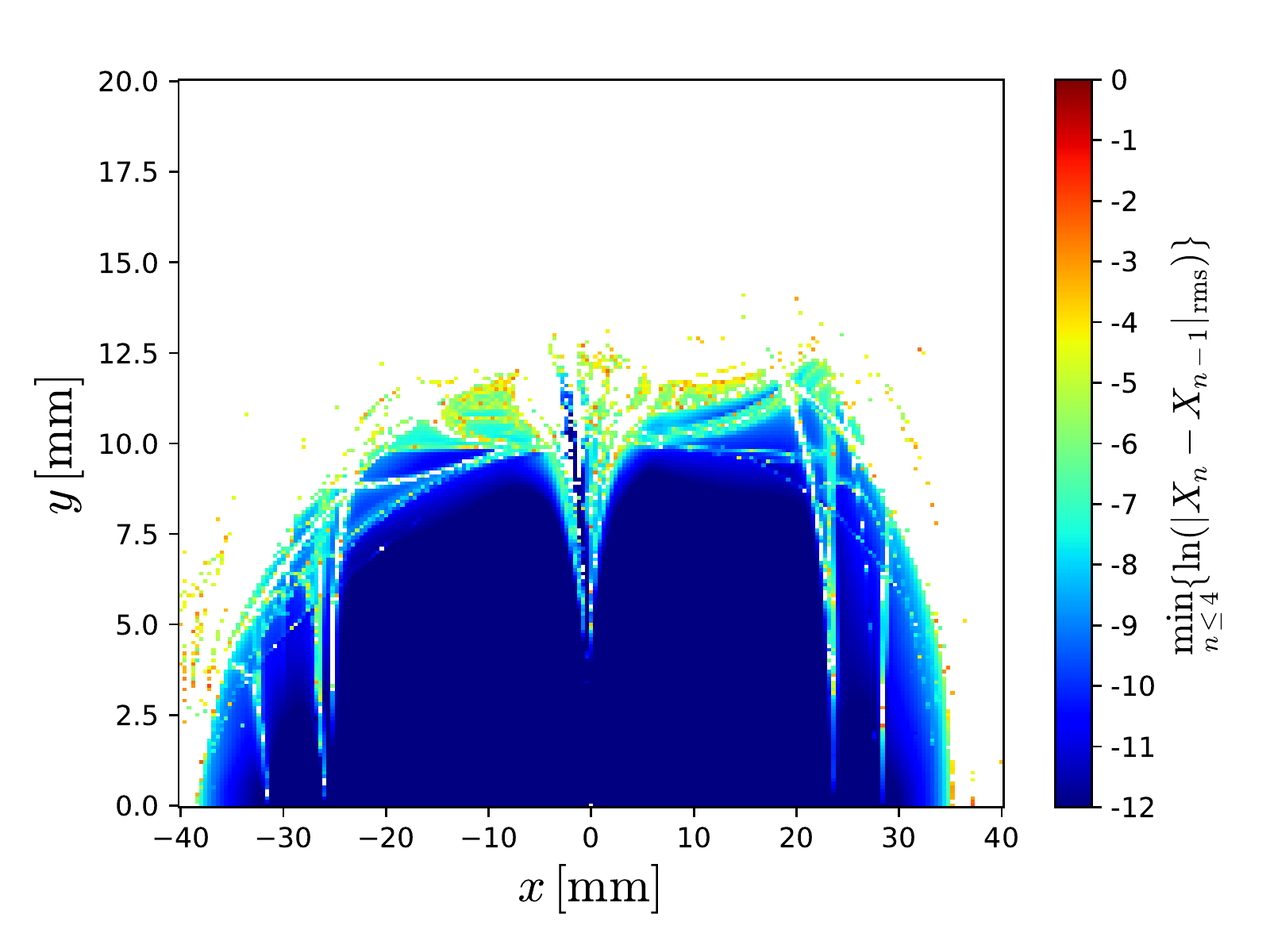}\includegraphics[width=0.45\textwidth]{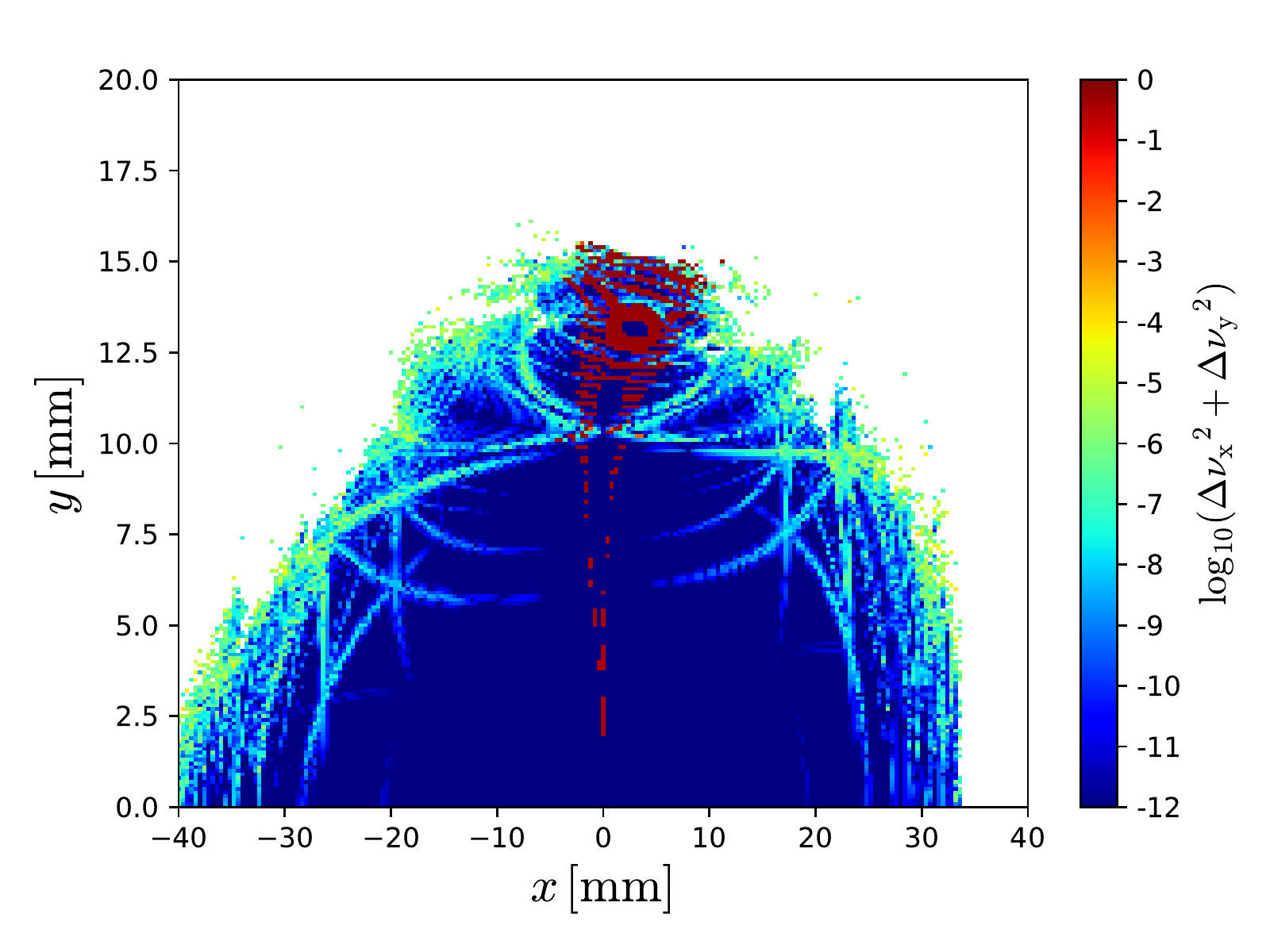}

\caption{The convergence map (left figure) and the frequency map (right figure) for NSLS-II bare lattice.}

\end{figure}

\begin{figure}[!htbp]
\includegraphics[width=0.45\textwidth]{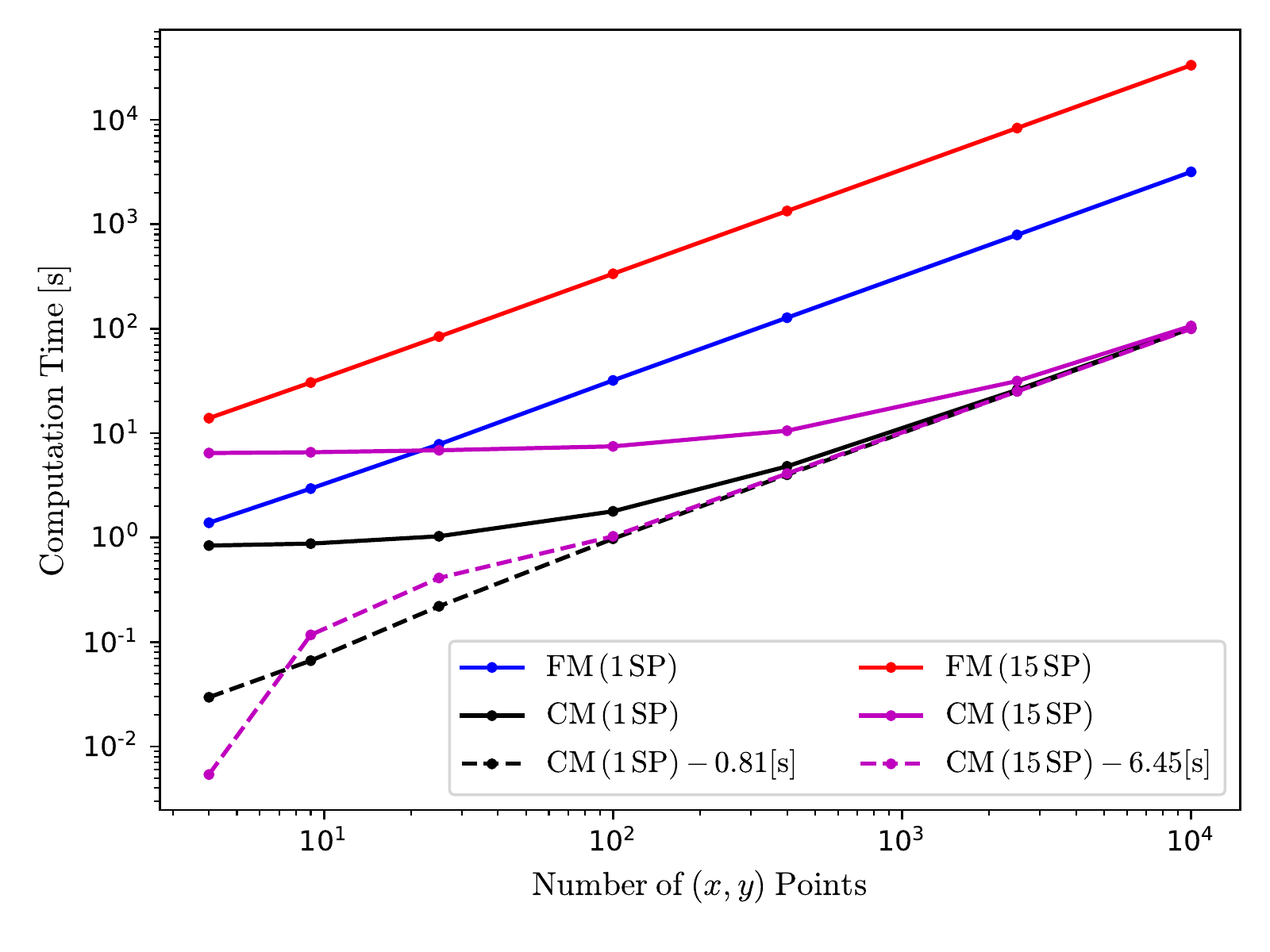}
\caption{Computation time comparison between convergence map (CM) and frequency
map (FM) analyses with different numbers of initial transverse coordinate
points and with one super-period (SP) and the whole ring (15 SPs)
of NSLS-II. The dash lines correspond to the computation times without
the initial setup times (mainly TPSA calculations). }

\end{figure}

In Section II.A, we first introduce the square matrix equation for
nonlinear dynamics, using the Henon map \cite{wayne,hao}
as an example. Then, in Section II.B, we show that a set of polynomials
derived from the left eigenvectors in the Jordan decomposition of
the square matrix can be used as a set of approximate action-angle
variables, i.e. the trajectory represented by these variables follow
a circle with a small deviation from a rotation with a constant rotation
speed. 

In Section III, we show that a suitable linear combination of these
action-angle variables can be used to minimize the deviation from
a pure rotation (i.e., a deviation from an exact action-angle variable).
Then, considering the small deviation as a perturbation, we show it
is possible to write an exact equation for the action-angle variables
with the deviation as a perturbation term. We develop a method to
solve this perturbation equation of motion using Fourier transform.
That is, we develop an iteration procedure to generate a sequence
of new action-angle variables to improve their precision, so in each
iteration step the new solution has not only less deviation from a
pure rotation, it is also closer to the exact solution of the equation.
In short, we use an analytical periodic function to approximate the
exact solution. When the iteration process is convergent the sequence
approaches the exact solution.

In Section IV, we test the precision numerically using the convergence
of the sequence as a criterion for the deviation of the solution from
a pure rotation. Actually, the convergence criterion can be used to
clarify the meaning of a ``small''
deviation from a pure rotation: if the deviation is ``small'',
the sequence should be convergent. Applying the convergence criterion
in the x,y plane leads to a ``convergence map'' ,
which is entirely different from the frequency map \cite{laskar},
but carries similar information about the trajectory, amplitude-tune
dependence, stability of the trajectory and the dynamic aperture. 

In Section V, we describe the convergence map application to nonlinear
lattice optimization using NSLS-II as an example, in comparison with
particle tracking.

In Section VI, we show the convergence map is much faster than tracking
for the same number of points taken for the phase space. Hence the
convergence map can be used as an efficient tool for the nonlinear
optimization of storage ring lattices. 

Then in Section VII, we compare the particle survival turn numbers
in tracking with the number of points taken for the action-angle variables
in a period of the trajectory before the iteration procedure diverges.
Our numerical study indicates their similar relation to dynamic aperture.
This is the reason why the convergence map can be used to study the
dynamic aperture. 

Section VIII is the conclusion.

\section{Square matrix}

\subsection{Square matrix equation for nonlinear dynamics}

We consider the equations of motion of a nonlinear dynamic system
with periodic structure such as Hills equation, it can be expressed
by a square matrix.

If we use the complex Courant-Snyder variable $z=x-ip$, its conjugate
and powers $z,z^{\ast},z^{2},...$ as a column vector $Z$, the one turn
map can be represented by a large square matrix $M$ using
\begin{equation}
Z=MZ_{0}
\end{equation}

The column vector $Z_{0}$ represents the initial value of the column vector $Z$
before the one turn mapping.

For an example of H\'enon map \cite{wayne}, 
\begin{align}\label{eq:henonmap_xp}
\begin{split}
x&=x_{0}\cos\mu+p_{0}\sin\mu+\epsilon x_{0}^{2}\sin\mu\\
p&=-x_{0}\sin\mu+p_{0}\cos\mu+\epsilon x_{0}^{2}\cos\mu
\end{split}
,
\end{align}
we use a variable transformation $z=x-ip$ , $z^{\ast}=x+ip$ ,$z_{0}=x_{0}-ip_{0}$
and $z_{0}^{\ast}=x_{0}+ip_{0}$ to rewrite this equation into a form
of the first two rows of the following equation Eq.(\ref{eq:henon_z}).
Then, using these firt two rows, $z,z^{\ast}$ and their higher power
monomials after one turn of rotation, or after one element in an accelerator
lattice, can be written as a truncated power series expansion of the
initial $z_{0}=x_{0}-ip_{0}$ and $z_{0}^{\ast}=x_{0}+ip_{0}$. For
example, up to 3rd order, we have: 
\begin{equation}
\begin{split}  z&=e^{i\mu}z_{0}-\frac{i}{4}\epsilon e^{i\mu}z_{0}^{2}-\frac{i}{2}\epsilon e^{i\mu}z_{0}z_{0}^{\ast}-\frac{i}{4}\epsilon e^{i\mu}z_{0}^{\ast2}\\
  z^{\ast}&=e^{-i\mu}z_{0}^{\ast}+\frac{i}{4}\epsilon e^{-i\mu}z_{0}^{2}+\frac{i}{2}\epsilon e^{-i\mu}z_{0}z_{0}^{\ast}+\frac{i}{4}\epsilon e^{-i\mu}z_{0}^{\ast2}\\
  z^{2}&=e^{2i\mu}z_{0}^{2}-\frac{i}{2}\epsilon e^{2i\mu}z_{0}^{3}-i\epsilon e^{2i\mu}z_{0}^{2}z_{0}^{\ast}-\frac{i}{2}\epsilon e^{2i\mu}z_{0}z_{0}^{\ast2}\\
  zz^{\ast}&=z_{0}z_{0}^{\ast}+\frac{i}{4}\epsilon z_{0}^{3}+\frac{i}{4}\epsilon z_{0}^{2}z_{0}^{\ast}-\frac{i}{4}\epsilon z_{0}z_{0}^{\ast2}-\frac{i}{4}\epsilon z_{0}^{\ast3}\\
  z^{\ast2}&=e^{-2i\mu}z_{0}^{\ast2}+\frac{i\epsilon}{2}e^{-2i\mu}z_{0}^{2}z_{0}^{\ast}+i\epsilon e^{-2i\mu}z_{0}z_{0}^{\ast2}+\frac{i\epsilon}{2}e^{-2i\mu}z_{0}^{\ast3}\\
  z^{3}&=e^{3i\mu}z_{0}^{3}\\
  ...&\\
  z^{\ast3}&=e^{-3i\mu}z_{0}^{\ast3}
\end{split}
,\text{ with \hspace{10bp}}Z\equiv\left(\begin{array}{c}
z\\
z^{\ast}\\
z^{2}\\
zz^{\ast}\\
z^{\ast2}\\
z^{3}\\
\vdots\\
z^{\ast3}
\end{array}\right)\label{eq:henon_z}
\end{equation}

Because in this equation the monomial term of power
order $m$ at the left hand side only have power terms of initial
monomial terms with power order higher than $m$ at the right hand
side, the coefficients in Eq.\eqref{eq:henon_z} form an upper-triangular
$10\times10$ square matrix $M,$ such that Eq.\eqref{eq:henon_z}
can be written as $Z=MZ_{0}$. In general, there are constant terms
in the expansion. In this example, the offset of x is zero, so the
constant terms are also zeros. The vector $Z$ spans a 10 dimensional linear space. The matrix $M$, when operated on $Z_{0}$ , represents
a rotation of $Z$ in this space. We remark here that even though
we mostly use $M$ to represent one turn map for a storage ring, each element in the storage ring dynamics or other nonlinear dynamics problem can also be written as a square matrix, then $M$ would be a product of the square matrix of the elements.

\subsection{Eigenvectors of Jordan blocks of square matrix $M$ as Approximate Action-Angle Variables}

All square matrices can be transformed into Jordan form \cite{Kagstrom1, Kagstrom2},
this transform is particularly very simple for a triangular square
matrix. A detailed description is given, e.g., in Ref \cite{yu2,Kagstrom1, Kagstrom2}.
For any given square matrix $M$, there are well known methods to
calculate an eigenvalue $\mu$, a transformation matrix $U$ and a
Jordan matrix $\tau$ so that every row of the matrix $U$ is a (generalized)
left eigenvector of $M$, with the i$^{\text{th}}$ row of $U$ denoted by $u_{i}$
satisfying 
\begin{equation}
\begin{split}UM=e^{i\mu I+{\tau}}U\end{split}
\label{eq:umuinv}
\end{equation}

As an example, for the case of the Henon map in Eq.\ref{eq:henonmap_xp} and Eq. \ref{eq:henon_z} with tune $\mu$ , one of the eigenvalues of $M$ is $e^{i\mu}$, the Jordan matrix $\tau$ has the form 
\begin{equation}\label{eq:tau2by2}
\tau=\begin{bmatrix}0 & 1\\
0 & 0
\end{bmatrix}  
\end{equation}
, with $I$ as the identity matrix. The Matrix U can be found as

\begin{align}
U=\left(\begin{array}{c}
u_{0}\\
u_{1}
\end{array}\right)
=\left({\scriptscriptstyle \begin{array}{cccccccccc}
0 & 1 & 0 & \frac{i}{4\left(-1+e^{i\mu}\right)} & \frac{ie^{i\mu}}{2-2e^{i\mu}} & \frac{ie^{3i\mu}}{4-4e^{3i\mu}} & -\frac{1}{8\left(-1+e^{i\mu}\right)^{2}} & 0 & U_{0,8} & U_{0,9}\\
0 & 0 & 0 & 0 & 0 & 0 & 0 & U_{1,7} & 0 & 0
\end{array}}\right)\label{eq:Uhenon}
\end{align}

with 
\begin{align*}
U_{0,8}&=\frac{e^{2i\mu}-e^{3i\mu}+e^{4i\mu}}{-8e^{i\mu}-8e^{3i\mu}+8e^{4i\mu}+8}\\
U_{0,9}&=-\frac{e^{5i\mu}}{8\left(-1+e^{i\mu}\right)^{2}\left(1+e^{2i\mu}\right)\left(e^{i\mu}+e^{2i\mu}+1\right)}\\
U_{1,7}&=\frac{e^{i\mu}\left(3e^{i\mu}+3e^{2i\mu}+2e^{3i\mu}+2\right)}{8\left(-1+e^{3i\mu}\right)}
\end{align*}

In the general case, the Jordan matrix $\tau$ always has much lower
dimension than the mapping matrix $M$, and has the form 
\begin{equation}
\tau=\begin{bmatrix}0 & 1 & 0 & ... & 0\\
0 & 0 & 1 & ... & 0\\
0 & 0 & ... & ... & 0\\
0 & 0 & 0 & ... & 1\\
0 & 0 & 0 & ... & 0
\end{bmatrix}.\label{eq:tau_nd}
\end{equation}

In the example for the case of 4 variables $x,p_{x},y,p_{y}$ at $3^\text{rd}$
order, as for the storage ring lattice example to be used later, the
matrix $M$ is a $35\times35$ matrix, Jordan matrix
$\tau$ is exactly same as the form of Eq.\ref{eq:tau2by2}, the matrices $U_{x}$ and $U_{y}$ are $2\times35$ transformation
matrix, for eigenvalues $\mu_{x}$ and $\mu_{y}$ respectively. For the convergence
map study in our example, high precision is achieved without using
power order higher than 3, and when it is convergent the result approaches the solution precisely.

As $Z=MZ_{0}$, Eq.\eqref{eq:umuinv}
gives 
\begin{equation}
UZ=UMZ_{0}=e^{i\mu I+{\tau}}UZ_{0}.\label{eq:uz_umz_0}
\end{equation}

Therefore a transformation is defined as 
\begin{equation}
\begin{split}W & \equiv UZ\\
W_{0} & \equiv UZ_{0}
\end{split}
\label{eq:w_transformation}
\end{equation}

$W$ represents the projection of the vector $Z$ onto the invariant
subspace spanned by the left eigenvectors $u_{j}$ given by the rows of the matrix $U$, such that each row of $W$ is $w_{j}=u_{j}Z$,
a polynomial of $z,z^{\ast}$. Then Eq.\eqref{eq:uz_umz_0} implies
the operation of one turn map $Z=MZ_{0}$, corresponds to a rotation
in the invariant subspace represented by 
\begin{equation}
W=e^{i\mu I+{\tau}}W_{0}.\label{eq:Wrot}
\end{equation}

As an example, for the Henon map Eq.(\ref{eq:henonmap_xp},\ref{eq:henon_z}),
because $U$ and $Z$ are given by Eq.(\ref{eq:Uhenon}, \ref{eq:henon_z})
we have:

\begin{align*}
w_{0}&=u_{0}Z=&z+\frac{i}{4\left(-1+e^{i\mu}\right)}z^{2}+\frac{ie^{i\mu}}{2-2e^{i\mu}}zz^{\ast}+\frac{ie^{3i\mu}}{4-4e^{3i\mu}}z^{\ast2}\\
&&-\frac{1}{8\left(-1+e^{i\mu}\right)^{2}}z^{3}
+\frac{e^{2i\mu}-e^{3i\mu}+e^{4i\mu}}{-8e^{i\mu}-8e^{3i\mu}+8e^{4i\mu}+8}zz^{\ast2}\\
&&-\frac{e^{5i\mu}}{8\left(-1+e^{i\mu}\right)^{2}\left(1+e^{2i\mu}\right)\left(e^{i\mu}+e^{2i\mu}+1\right)}z^{\ast3}\\
w_{1}&=u_{1}Z=&\frac{e^{i\mu}\left(3e^{i\mu}+3e^{2i\mu}+2e^{3i\mu}+2\right)}{8\left(-1+e^{3i\mu}\right)}z^{2}z^{\ast}
\end{align*}

The lowest power order term of $w_{0}$ is the linear
term $z$, compared with 3rd power order term $z^{2}z^{\ast}$ for $w_1$. 

In general, for small amplitude the lowest order terms in $W$ dominate.
Since $W$ rotates in the invariant subspace of $Z$, for sufficiently
small amplitude, when the higher power term (for example in H\'enon example, Eq.\eqref{eq:henon_z})
is negligibly small, the equation is nearly linear, hence the absolute
value of each row $|w_{j}|=|u_{j}Z|$ of $W$ is approximately invariant
with a phase advance given by $\mu+\phi$, where $\mu$ is the linear
tune, and $\phi\ll\mu$ is the amplitude dependent tune shift. This is related to the
KAM theory in nonlinear dynamics.

KAM theory states that the invariant tori are stable under small perturbation
(See, for example,  Ref \cite{lieberman, broer,arnold}).
In our examples, for sufficiently small amplitude of oscillation in $z$, the invariant tori are deformed and survive, i.e., the motion is quasiperiodic. So the system has a nearly stable frequency, and when the amplitude is small, the fluctuation of the frequency is also small. Since during the dynamical process $W$ remains in the eigenspace of the column space $Z$, after $n$ turns, approximately, the vector $W$ only changes by a phase factor $e^{in(\mu+\phi)}$, i.e.,

\begin{equation}
W=e^{n(i\mu I+{\tau})}W_{0}\cong e^{in(\mu+\phi)}W_{0}.\label{eq:wrotate}
\end{equation}
From a comparison of both sides of Eq.\eqref{eq:wrotate} we have,
\begin{equation}
\tau W_{0}\cong i\phi W_{0}.\label{eq:weigenvec}
\end{equation}
and $\phi$ is the amplitude dependent tune shift. In Eq.\eqref{eq:weigenvec} we use the approximate equal sign because for a matrix $\tau$ of
finite dimension m, the relation is only an approximation for sufficiently
small amplitude. We write the Eq. \eqref{eq:wrotate} explicitly
using the property of the Jordan matrix $\tau$ given by Eq. \eqref{eq:tau_nd}
as a raising operator: 
\begin{equation}
\tau\begin{bmatrix}w_{0}\\
w_{1}\\
...\\
w_{m-1}
\end{bmatrix}=\begin{bmatrix}w_{1}\\
w_{2}\\
...\\
0
\end{bmatrix}\cong\begin{bmatrix}i\phi w_{0}\\
i\phi w_{1}\\
...\\
i\phi w_{m-1}
\end{bmatrix}\label{eq:wshift}
\end{equation}
where $w_{j}$'s are the rows of $W_{0}$. Compare the two sides we
find 
\begin{equation}
i\phi=\frac{w_{1}}{w_{0}}\cong\frac{w_{2}}{w_{1}}\cong\frac{w_{3}}{w_{2}}\dots\label{eq:tuneshift}
\end{equation}

In the study of truncated power series, m is finite, hence Eq. \eqref{eq:tuneshift}
is an approximation. The polynomials $w_{m-2},w_{m-1}$ have only
high order terms, and as $m$ increases, when the amplitude of $z$
is sufficiently small, the last term in Eq. \eqref{eq:tuneshift}
becomes the ratio of two negligibly small numbers, and is less accurate.
Actually the last row of Eq. \eqref{eq:wshift} is impossible, so
it can only be taken as an approximation representing the fact that
$|\phi w_{m-1}|$ is neglegibly small. In addition to the condition
Eq. \eqref{eq:tuneshift} for a stable motion, obviously, for the
amplitude to be nearly constant, another condition is 
\begin{equation}
\text{Im}(\phi)\cong0.\label{eq:imphizero}
\end{equation}

Fig.3 compare the tracking (direct iteration of Eq.\eqref{eq:henonmap_xp},
red) and $\nu=(\mu+\phi)/2\pi$ (green), with $i\phi=\frac{w_{1}}{w_{0}}$
of Eq.(\ref{eq:tuneshift}). It is clear that $w_{0}$ can be used as
an aproximate action-angle variable, even near the resonances $x\sim(0.53,0.68)$.

\begin{figure}[!htbp]
\includegraphics[width=0.9\columnwidth]{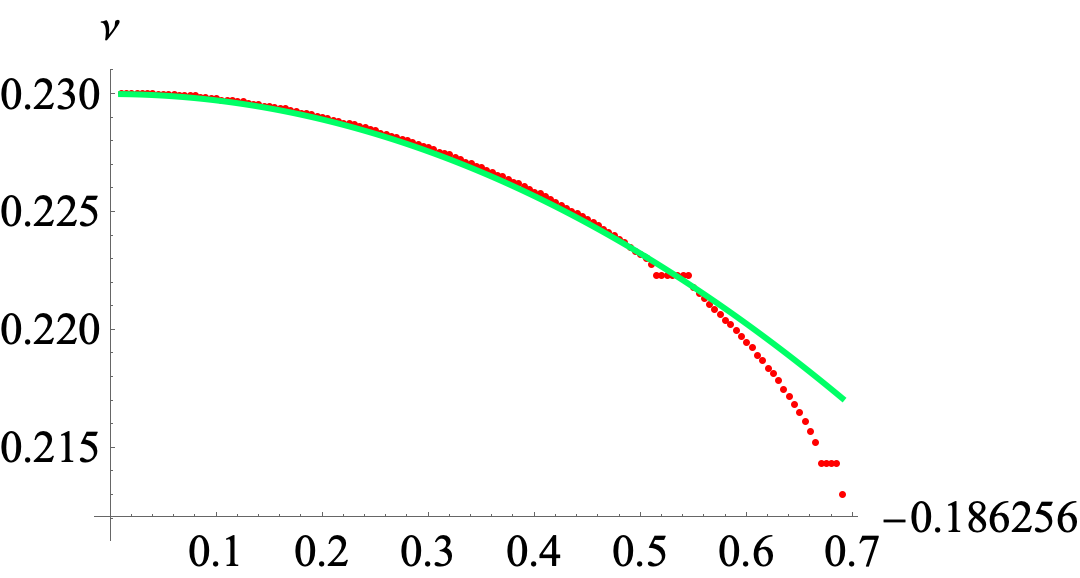}

\caption{Compare tune $\nu$ vs.$x$ calculated by tracking with square matrix }

\end{figure}

In our study of storage ring lattice using its matrix $M$, Eqs.\eqref{eq:tuneshift} and \eqref{eq:imphizero}
are confirmed by many numerical examples.

The Eq.\eqref{eq:wrotate}, derived for $x,y$ planes separately,
leads to a set of polynomials $w_{x0},w_{x1},\cdots$, and $w_{y0},w_{y1},\cdots$.
Our tracking results confirmed that these polynomials, or linear combinations
can be used as a set of approximate action-angle variables: after
$n$ turns they are multiplied by a factor of form $e^{in(\mu+\phi)}$
approximately, ie. the trajectory represented by these variabes follow
a circle with small deviation from a rotation with uniform rotation
speed. In addition, we find that near the stability border the deviation
of these actions from constancy provides a measure of the destruction
of invariant tori, or a measure of the stability of trajectories and
tunes. In the next section, we shall  seek for a variable transformation
from $x,p_{x},y,p_{y}$ to these action-angle variables such that the nonlinear dynamical equation is transformed into 
a form which can be solved by an iteration method . 

\section{\label{sec:Perturbation}Perturbation based on action-angle approximation and iteration}

In this section, we will detail the iteration method to find the action-angle approximation and its application in predicting particle's long term stability.
\subsection{Variable transform of dynamical quation and perturbation solution}

In the previous section, a transformation $U$ is generated by the square
matrix method for nonlinear map $M$ to create the new set of variables
$w$, so that map $U\circ M\circ U^{-1}$ is approximately a rigid
rotation and $w$ serve as approximate action-angle variables. Since
the new set of variable $w$ is not unique, we use $v_{1}$ and $v_{2}$
to denote the choice of approximate action. The simplest choice
is $v_{1}=w_{x0}$ and $v_{2}=w_{y0}$. Further variable transformation can be found to make the map to be an exact rigid rotation map when there is a quasi-periodic solution for the system.

To formulate the perturbation problem, we rewrite the action-angle
variables $v_{1},v_{2}$ as 

\begin{align}\label{eq:vdef}
\begin{split}
 & v_{1}(\theta_{1})\equiv e^{i\theta_{1}}\\
 & v_{2}(\theta_{2})\equiv e^{i\theta_{2}}
 \end{split}
\end{align}

where $\theta_{1},\theta_{2}$ are complex numbers which denote oscillations
deviate from rigid rotation with unknown frequency $\omega_{1},\omega_{2}$
with small phase fluctuation (the real part of $\theta_{1},\theta_{2}$
) and amplitude fluctuation (imaginary part of $\theta_{1},\theta_{2}$).
We consider the relation between the 4 variables, $\theta_{1},\theta_{2}$
, their complex conjugate $\theta_{1}^{*},\theta_{2}^{*}$, and $x,p_{x},y,p_{y}$
(to be abbreviated as $X$ in the following) as a variable transformation.

We establish one turn map using $\theta_{1},\theta_{2}$ as dynamic
variables, i.e, we consider $\theta_{1,k+1},\theta_{2,k+1}$ as function
of $\theta_{1k},\theta_{2k}$ (with $k$ the turn number):

\begin{align}\label{eq:vdef_phi}
\begin{split}
 & \theta_{1,k+1}=\theta_{1,k}-i\ log\frac{v_{1}(\theta_{1,k+1})}{v_{1}(\theta_{1,k})}\equiv\theta_{1,k}+\phi_{1}(\theta_{1,k},\theta_{2,k})\\
 & \theta_{2,k+1}=\theta_{2,k}-i\ log\frac{v_{2}(\theta_{2,k+1})}{v_{2}(\theta_{2,k})}\equiv\theta_{2,k}+\phi_{2}(\theta_{1,k},\theta_{2,k}) 
\end{split}
\end{align}
where $\phi_{1}$ and $\phi_{2}$ are functions which can be found
from the map without approximations. Now when there are quasi-periodic
solutions we can find a transformation from $(\theta_{1},\theta_{2})$
to a rigid rotation variable pair $(\alpha_{1},\alpha_{2})$, which
satisfy

\begin{align}\label{eq:rigid_rotation}\begin{split}
 & \alpha_{1,k+1}=\alpha_{1,k}+\omega_{1}\\
 & \alpha_{2,k+1}=\alpha_{2,k}+\omega_{2} 
\end{split}
\end{align}
where $\omega_{1}$ and $\omega_{2}$ are the rotation number of the
map. By denoting $\Delta\phi_{1,2}\equiv\phi_{1,2}-\omega_{1,2}$,
the equations for $(\theta_{1},\theta_{2})$ become

\begin{align}\label{eq:dynamics_theta}
\begin{split}
 & \theta_{1}(\alpha_{1}+\omega_{1},\alpha_{2}+\omega_{2})-\theta_{1}(\alpha_{1},\alpha_{2})=\omega_{1}+\Delta\phi_{1}(\theta_{1}(\alpha_{1},\alpha_{2}),\theta_{2}(\alpha_{1},\alpha_{2}))\\
 & \theta_{2}(\alpha_{1}+\omega_{1},\alpha_{2}+\omega_{2})-\theta_{2}(\alpha_{1},\alpha_{2})=\omega_{2}+\Delta\phi_{2}(\theta_{1}(\alpha_{1},\alpha_{2}),\theta_{2}(\alpha_{1},\alpha_{2}))
 \end{split}
\end{align}

If $v_{1},v_{2}$ are sufficiently close to a pure rotation, then
$|\Delta\phi_{1}|,|\Delta\phi_{2}|\ll1$. The real part of $|\Delta\phi_{1}|,|\Delta\phi_{2}|$
are the phase fluctuation, and their imaginary part is the amplitude
fluctuation. This exact equation can be solved by perturbation: initially
we take zero order approximation $\theta_{1}^{(0)}=\alpha_{1},\theta_{2}^{(0)}=\alpha_{2}$.
Then $X^{(0)}$ is calculated by the inverse function of Eq.\eqref{eq:vdef},
and $\Delta\phi_{1}^{(0)},\Delta\phi_{1}^{(0)}$is calculated as given
by Eq. \eqref{eq:vdef_phi}. When this is substituted to the right hand
side of Eq. \eqref{eq:dynamics_theta}, since $|\Delta\phi_{1}|,|\Delta\phi_{2}|\ll1$
the error is of second order. Hence to first order the solution satisfies 

\begin{align}\label{eq:dynamics_theta_0}
\begin{split}
 & \theta_{1}^{(1)}(\alpha_{1}+\omega_{1},\alpha_{2}+\omega_{2})-\theta_{1}^{(1)}(\alpha_{1},\alpha_{2})\approx\omega_{1}^{(0)}+\Delta\phi_{1}(\theta_{1}^{(0)}(\alpha_{1},\alpha_{2}),\theta_{2}^{(0)}(\alpha_{1},\alpha_{2})) \\
 & \theta_{2}^{(1)}(\alpha_{1}+\omega_{1},\alpha_{2}+\omega_{2})-\theta_{2}^{(1)}(\alpha_{1},\alpha_{2})\approx\omega_{2}^{(0)}+\Delta\phi_{2}(\theta_{1}^{(0)}(\alpha_{1},\alpha_{2}),\theta_{2}^{(0)}(\alpha_{1},\alpha_{2}))
 \end{split}
\end{align}
where $\omega_{1}^{(1)},\omega_{2}^{(1)}$ are the constant term of
the Fourier transform of $\phi_{1}(\alpha_{1},\alpha_{2}),\phi_{2}(\alpha_{1},\alpha_{2})$.
This equation can be solved using Fourier transform (see Appendix
A about the solution by Fourier transform), the result is the first
order approximation $\theta_{1}^{(1)},\theta_{2}^{(1)}$ with frequency
$\omega_{1}^{(1)},\omega_{2}^{(1)}$. Then $\theta_{1}^{(1)},\theta_{2}^{(1)}$
is substituted in the right hand side of Eq. \eqref{eq:dynamics_theta_0} 
to obtain $\theta_{1}^{(2)},\theta_{2}^{(2)}$ as $2^\text{nd}$ order approximation. 

\subsection{Iteration of perturbation solution}

This process can be iterated to generate convergent solution to high
precision if the amplitude of zero'th order$X^{(0)}$ (obtained from
$\theta_{1}^{(0)},\theta_{2}^{(0)}$) is sufficiently close the origin,
within the dynamical aperture. At iteration step $k$ (i.e., after
k iteration ), the equation is

\begin{align}\label{eq:dynamics_theta_k}
\begin{split}
 & \theta_{1}^{(k+1)}(\alpha_{1}+\omega_{1},\alpha_{2}+\omega_{2})-\theta_{1}^{(k+1)}(\alpha_{1},\alpha_{2})\approx\omega_{1}^{(k)}+\Delta\phi_{1}^{(k)}(\alpha_{1},\alpha_{2}) \\
 & \theta_{2}^{(k+1)}(\alpha_{1}+\omega_{1},\alpha_{2}+\omega_{2})-\theta_{2}^{(k+1)}(\alpha_{1},\alpha_{2})\approx\omega_{2}^{(k)}+\Delta\phi_{2}^{(k)}(\alpha_{1},\alpha_{2})
 \end{split}
\end{align}
where the subscripts denote the iteration number of the corresponding variables. $\omega_{1}^{(k)},\omega_{2}^{(k)}$ are the constant terms of the
Fourier transform of $\phi_{1}^{(k)}(\alpha_{1},\alpha_{2}),\phi_{2}^{(k)}(\alpha_{1},\alpha_{2})$.
For solution $\theta_{1}^{(k)}(\alpha_{1},\alpha_{2}),\theta_{2}^{(k)}(\alpha_{1},\alpha_{2})$
the corresponding coordinates, denoted as $X^{(k)}$, are calculated
as the inverse function of Eq.\eqref{eq:vdef}.

The main issue is the convergence of this iteration process. In the
neighbourhood of a pure rotation, KAM theory \cite{lieberman,arnold,broer}
proved the existence of analytical solution. In a practical application,
instead of trying to prove the existence of exact analytical solution,
we apply iteration procedure to find the quasi-periodic solution exploring
area with large amplitude or near resonance numerically.

\subsection{Minimize deviation from pure rotation by renewing linear combination
coefficients within an iteration step}

In the numerical tests, we found that keeping the simplest choice, i.e.
$v_{1}=w_{x0}$ and $v_{2}=w_{y0}$ does not always yields a successful
iteration process for large amplitude particles when numerical simulation
suggests a quasi-static orbit exists. As shown in the last section,
the polynomials $w_{x0},w_{x1},\cdots$, and $w_{y0},w_{y1},\cdots$
may serve as approximate action angles, therefore we may extend the
choice of $v_{1}$and $v_{2}$ to be linear combinations of them to
allow the iteration method to start from a better approximation of
rigid rotation. The linear combination can be written as:

\begin{align}\label{eq:vdef_linear}
\begin{split}
& v_{1}(\theta_{1})\equiv a_{11}w_{x0}(X)+a_{12}w_{x1}(X)+a_{13}w_{y0}(X)+a_{14}w_{y1}(X)\\
& v_{2}(\theta_{2})\equiv a_{21}w_{x0}(X)+a_{22}w_{x1}(X)+a_{23}w_{y0}(X)+a_{24}w_{y1}(X)
\end{split}
\end{align}
with $a_{ij}$ as free parameters and $X$ denotes the dynamic variables
$(x,p_{x},y,p_{y})$. Here we use the 4 polynomials $w_{x0},w_{x1}$$,w_{y0},w_{y1}$,
because our experiences shows for 'convergence map' we found 4 is
often enough to generate the map. In some special cases, for example
when we calculate solution for some resonances, more polynomials such
as $w_{x2},w_{x3}$$,w_{y2},w_{y3}$ ,obtained from higher order Jordan
vectors are used to reach convergence.  But the study of solution
for resonances is still in progress, hence  here we limit our discussion
to only 4 polynomials. 

In Eq. \eqref{eq:dynamics_theta_0}, if the zeroth order approximation,
the pure rotation 
$v_{1}^{(0)}(\theta_{1}^{(0)}(\alpha_{1})),
v_{2}^{(0)}(\theta_{2}^{(0)}(\alpha_{2}))$
in Eq. \eqref{eq:vdef_linear}, determined by a set of linear combination
coefficients $\{a_{i,j}^{(0)}\},(i=1,2),(j=1,2,3,4\}$ (the choice
of initial $\{a_{ij}^{(0)}\}$ is discussed in the beginning of Appendix
A), are sufficiently close to the solution, then the perturbation
$|\Delta\phi_{1}^{(0)}|,|\Delta\phi_{2}^{(0)}|$ in Eq.\eqref{eq:dynamics_theta_0}
would be small, the Fourier expansion coefficients except the constant
terms ($\alpha_{1},\alpha_{2})$ in the Fourier expansion of $\theta_{1}^{(1)},\theta_{2}^{(1)}$
(see Appendix A) would be small: $|\widetilde{\theta^{(1)}}_{1nm}|\ll 1,\ |\widetilde{\theta^{(1)}}_{2nm}|\ll 1$.
The first order solution given by Eq. \eqref{eq:dynamics_theta_0} and the
first order approximation 
$v_{1}^{(0)}(\theta_{1}^{(1)}(\alpha_{1},\alpha_{2})),
v_{2}^{(0)}(\theta_{2}^{(1)}(\alpha_{1},\alpha_{2}))$
given by Eq.\eqref{eq:vdef_linear} would provide more accurate solution.
In order to enhance the convergence rate we add one step in the iteration,
i.e., we use a least square method to minimize the fluctuation term
$|\Delta\phi_{1}^{(1)}|,|\Delta\phi_{2}^{(1)}|$ in Eq. \eqref{eq:dynamics_theta_k}
by varying the linear combination $\{a_{i,j}\}$ in Eq. \eqref{eq:vdef_linear}.
The goal of this step is to minimize the deviation from pure rotation
of $v_{1}\equiv v_{1}(\theta_{1}^{(1)}(\alpha_{1},\alpha_{2})),$
$v_{2}\equiv v_{2}(\theta_{2}^{(1)}(\alpha_{1},\alpha_{2}))$ using
the optimized $\{a_{i,j}^{(1)}\}$ to generate $v_{1}^{(1)}\equiv v_{1}^{(1)}(\theta_{1}^{(1)}(\alpha_{1},\alpha_{2})),$
$v_{2}^{(1)}\equiv v_{2}^{(1)}(\theta_{2}^{(1)}(\alpha_{1},\alpha_{2}))$.
This least square method is described in Appendix A. 

In addition to this change of the linear combination coefficients
$\{a_{i,j}\}$ to make the orbit more close to a pure rotation, another
way to speed up the convergence is to further decouple the two sets
of points in the trajectory determined by either $\theta_{1}(\alpha_{1},\alpha_{2})$
with fixed $\alpha_{2}$, or by $\theta_{2}(\alpha_{1},\alpha_{2})$
with fixed $\alpha_{1}$. Since the Fourier transform of the trajectory
is given by $\widetilde{\theta}_{1nm}$ and $\widetilde{\theta}_{2nm}$
on $n_{\theta}^{2}$ points, as explained in Appendix A, the $\alpha_{1},\alpha_{2}$
phase space is also described by $n_{\theta}^{2}$ points determined
by $\alpha_{1i},\alpha_{2j}$, as shown in Fig.5a in the next example
section. When the solution is close to a pure rotation, there are
two nearly independent functions $\theta_{1}(\alpha_{1},\alpha_{2})\approx\theta_{1}(\alpha_{1})$
and $\theta_{2}(\alpha_{1},\alpha_{2})\approx\theta_{2}(\alpha_{2})$,
thus there are two nearly decoupled action-angle variables $v_{1},v_{2}$.
In the numerical examples in Section IV Fig.7c,d, we show the trajectory
in $v_{1},v_{2}$ planes separately as an example to see how they
are nearly decoupled even at the starting point $v_{1}^{(0)},v_{2}^{(0)}$,
as is more visible in Fig.7d. The blue dots (for varied $\alpha_{1})$
move aroud each red point with fixed $\alpha_{2}$ forming a small
circles. As iteration number increases, the circles reduced their
radius to points. This rapid decoupling is clear visible in Fig. 7a,b.
The convergence result agree with tracking very well, as will be explained
in Fig.6a of the next section. Hence for each fixed $\alpha_{2}$,
average $\theta_{2}$ over all $\alpha_{1}$ makes the points more
close to the poinst of a pure roation. Same way for each fixed $\alpha_{1}$,
we average $\theta_{1}$ over all $\alpha_{2}$. This speeds up the
decoupling of $v_{1},v_{2}$, and further speeds up the convergence
in our iterations. Thus the averaging process is included as part
of the second step.

This step is applied in every iteration $k$: Start from $\{a_{i,j}^{(k)}\}$,
$X^{(k)}(\alpha_{1},\alpha_{2})$, obtained in previous iteration,
we find $v_{1}^{(k)}(\alpha_{1},\alpha_{2}),v_{2}^{(k)}(\alpha_{1},\alpha_{2})$
and hence $\theta_{1}^{(k)}(\alpha_{1},\alpha_{2}),\theta_{2}^{(k)}(\alpha_{1},\alpha_{2})$
using Eq. \eqref{eq:vdef_linear}, then the solution of Eq. \eqref{eq:dynamics_theta_k}
gives $\theta_{1}^{(k+1)}(\alpha_{1},\alpha_{2}),\theta_{2}^{(k+1)}(\alpha_{1},\alpha_{2})$.
In turn, these lead to $v_{1}^{(k)}(\alpha_{1},\alpha_{2})\equiv v_{1}^{(k)}(\theta_{1}^{(k+1)}(\alpha_{1},\alpha_{2}))$
and $v_{2}^{(k)}(\alpha_{1},\alpha_{2})\equiv v_{2}^{(k)}(\theta_{2}^{(k+1)}(\alpha_{1},\alpha_{2}))$,
which can be used to find $X^{(k+1)}$ using the inverse function
of Eq. \eqref{eq:vdef_linear}. Then the least square method in the Appendix
B, followed by the decoupling averaging, is applied to find $\{a_{i,j}^{(k+1)}\}$.
Then, $\{a_{i,j}^{(k+1)}\}$ and $X^{(k+1)}$ are the staring point
of next iteration. The further optimized $\{a_{ij}^{(k+1)}\}$ in
Eq. \eqref{eq:vdef_linear} would correspond to a new set of 
$v_{1}^{(k+1)}(\alpha_{1},\alpha_{2})\equiv v_{1}^{(k+1)}(\theta_{1}^{(k+1)}(\alpha_{1},\alpha_{2}))$,
$v_{2}^{(k+1)}(\alpha_{1},\alpha_{2})\equiv v_{1}^{(k+1)}(\theta_{1}^{(k+1)}(\alpha_{1},\alpha_{2}))$
more close to a pure rotation. The two step cycle of iteration is
illustrated in Fig.4.

\begin{figure}[!htbp]
\includegraphics[width=0.7\columnwidth]{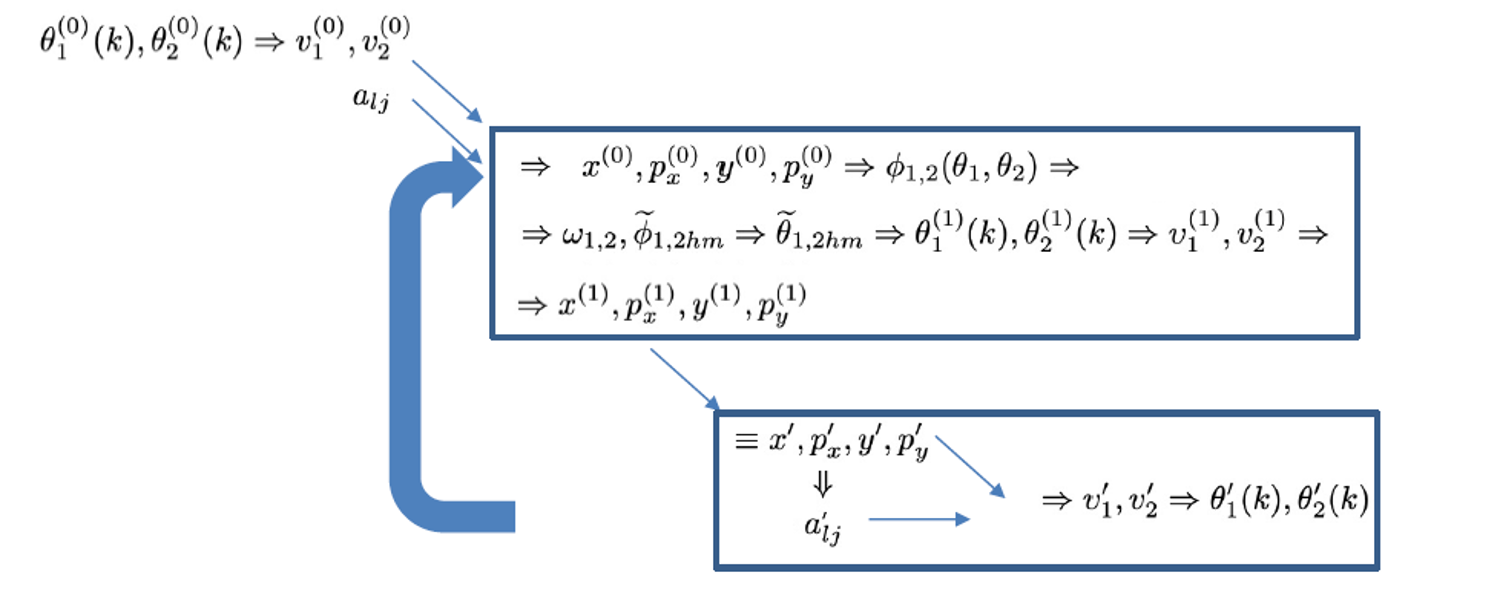}

\caption{flow diagram for iteration, where  $\widetilde{\theta} $, $\widetilde{\phi} $ are the Fourier trasform of $\theta $,$\phi $    }
\end{figure}

There are two steps here in every iteration $k$: 1. find $v_{1}^{(k)},v_{2}^{(k)}$
so it is more close to the exact solution, as the step given by Eq. \eqref{eq:dynamics_theta_k};
2. find new set of linear combination $\{a_{ij}^{(k+1)}\}$ so $v_{1}^{(k+1)},v_{2}^{(k+1)}$
is more close to a pure rotation, as given by Eq. \eqref{eq:vdef_linear}
using the least square method in Appendix A.

For each iteration step $k$, $v_{1}^{(k)}(\alpha_{1},\alpha_{2}),v_{2}^{(k)}(\alpha_{1},\alpha_{2})$
and corresponding linear combination $\{a_{ij}^{(k)}\}$ of polynomials
$w_{xj},w_{yj}$ give a periodic solution (trajectory) $X^{(k)}(\alpha_{1},\alpha_{2})\equiv x^{(k)}(\alpha_{1},\alpha_{2}),p_{x}^{(k)}(\alpha_{1},\alpha_{2}),y^{(k)}(\alpha_{1},\alpha_{2}),p_{y}^{(k)}(\alpha_{1},\alpha_{2})$.
We can use the convergence of $X^{(k)}(\alpha_{1},\alpha_{2})$ as
$k$ increases to test the convergence of the iteration based on the
Cauchy convergence criterion, i.e., we study $\delta X_{k}$, which
is the the standard deviation of $\delta X_{k}\equiv(X^{(k)}(\alpha_{1},\alpha_{2})-X^{(k-1)}(\alpha_{1},\alpha_{2}))_{rms}$,
and check whether the sequence of $\delta X_{k}$ decreases exponentially
as $k$ increases. If the iteration is convergent, as the iteration
$k$ increases, the solution approaches an accurate solution near a
limit determined by computer precision. Even though $v_{1}^{(k)},v_{2}^{(k)}$
are approximation to a pure rotation, generally they are not necessarily
approaching pure rotation as $k$ increases. 

Obviously the iteration method discussed in this section cannot be
applied to resonance region where the two action-angle variables becomes
correlated, and there is only one independent action-angle variable
left. The discussion about resonance case should be addressed in a
separate publication rather than this article.

With these provisions, we discuss the numerical application of this
iteration steps in the following. 

\section{Numerical application of convergence map }

In practical numerical application, one of the main parameters is
the number of indices $m,n={0,1,2,...n_{\theta-1}}$ in Eq. \eqref{eq29}.
Correspondingly in the inverse Fourier transform of Eq. \eqref{eq29}
the variables $\theta_{1},\theta_{2}$ and $\alpha_{1}$,$\alpha_{2}$
also take discrete values at $n_{\theta}^{2}$ points in the $\theta_{1},\theta_{2}$
plane. Fig.5 is an illustration of $\theta_{1}^{(0)}(\alpha),\theta_{2}^{(0)}(\alpha)$
and the result of tracking them one turn.

\begin{figure}[!htbp]
\includegraphics[width=0.5\columnwidth]{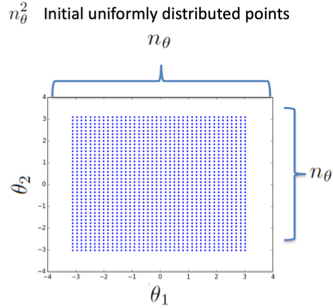}\includegraphics[width=0.5\columnwidth]{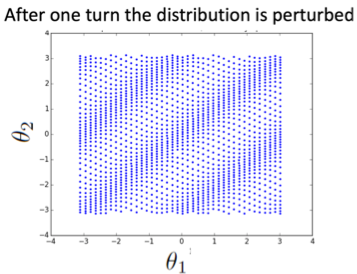}

\caption{points in initial $\theta_{1}^{(0)}(\alpha),\theta_{2}^{(0)}(\alpha)$,
and their distribution after one turn}

\end{figure}

\begin{figure}[!htbp]
\includegraphics[width=0.5\columnwidth]{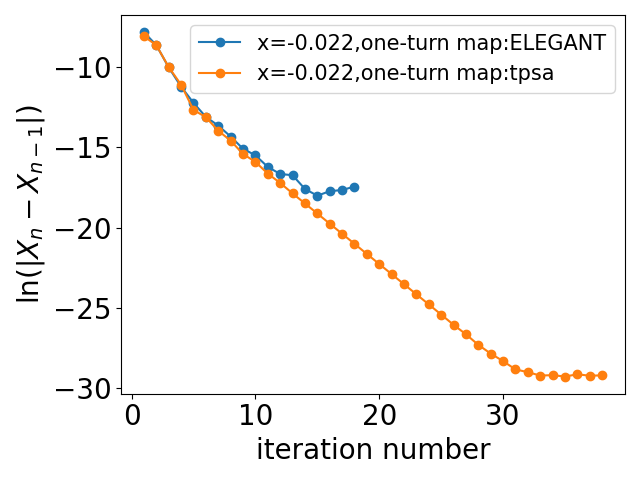}\includegraphics[width=0.5\columnwidth]{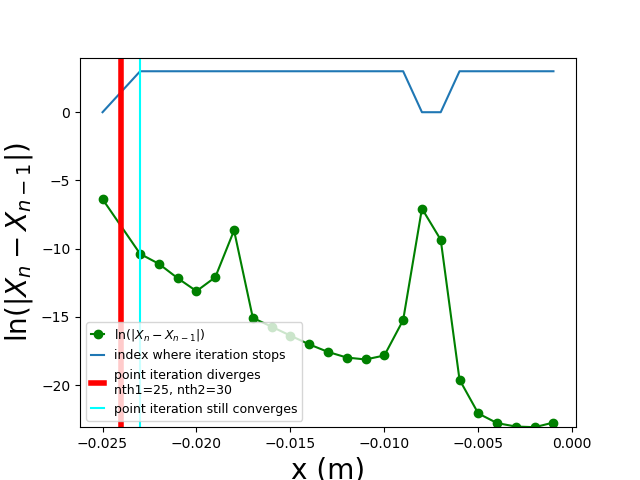}

\caption{(a) $\delta X_{n}$ vs. iteration number $n$; (b) Minimum
of $\delta X_{n}$ vs. $x$ , the red line is where iteration diverges
for both $n_{\theta}=$25 and 30 and the maximum number of iteration
is set at 4. }

\end{figure}

\begin{figure}[!htbp]
\includegraphics[width=0.5\columnwidth]{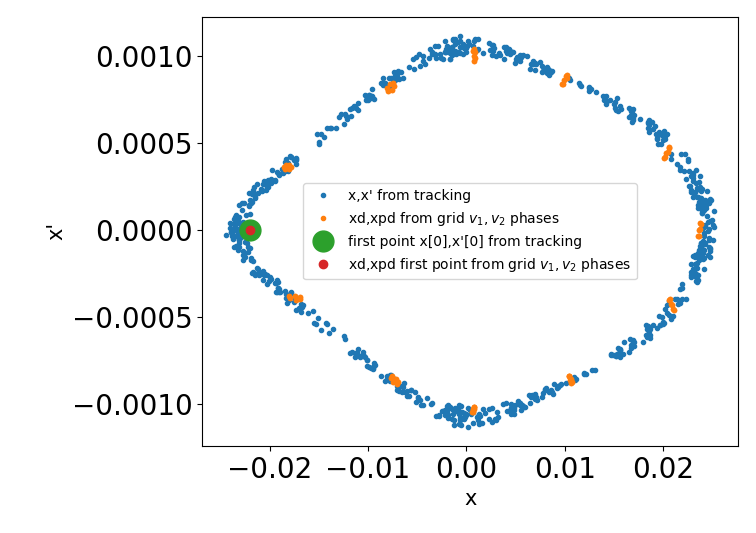}\includegraphics[width=0.5\columnwidth]{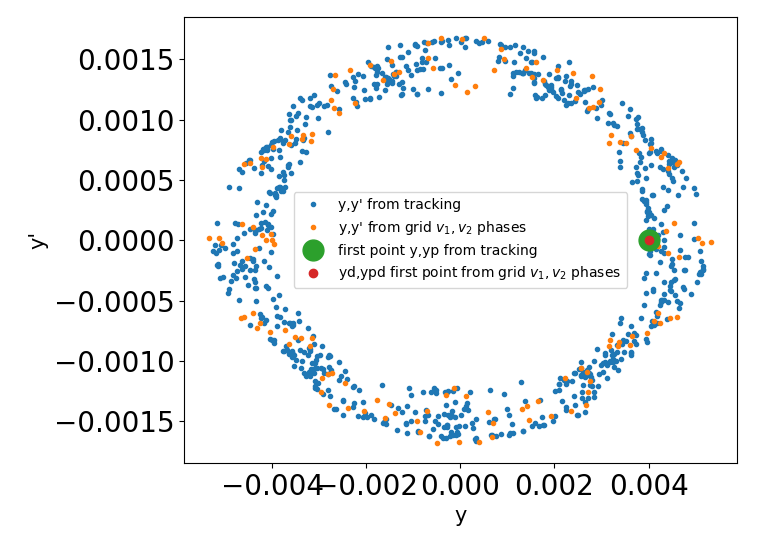}
\includegraphics[width=0.5\columnwidth]{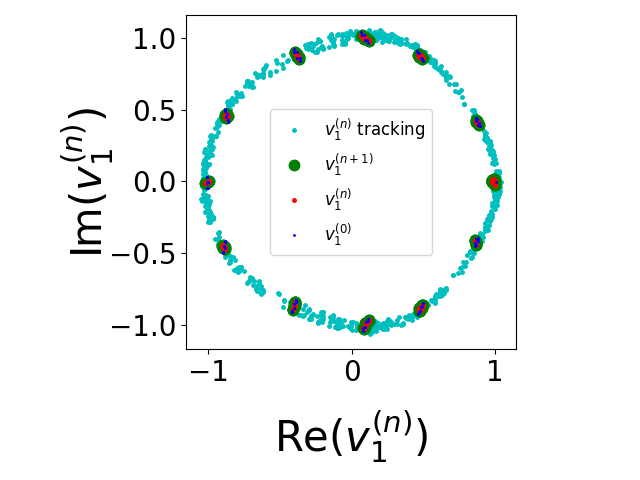}\includegraphics[width=0.5\columnwidth]{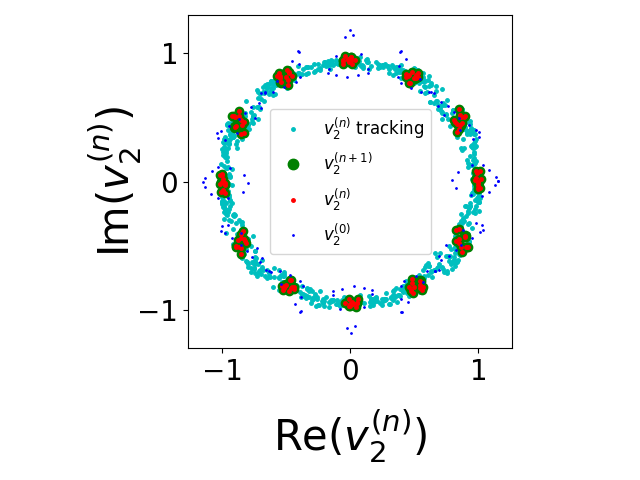}

\caption{(a) trajectory in $xp_{x}$ plane within the range $-3mm<y<3mm$.
\ (b) trajectory in $yp_{y}$ plane within the range $-2mm<x<2mm$,
blue (tracking by ELEGANT), orange (square
matrix iteration). (c) trajectory in $v_{1}$ plane, (d) trajectory
in $v_{2}$ plane. light blue (tracking by ELEGANT),
red (square matrix iteration)}

\end{figure}

\begin{figure}[!htbp]
\includegraphics[width=0.5\columnwidth]{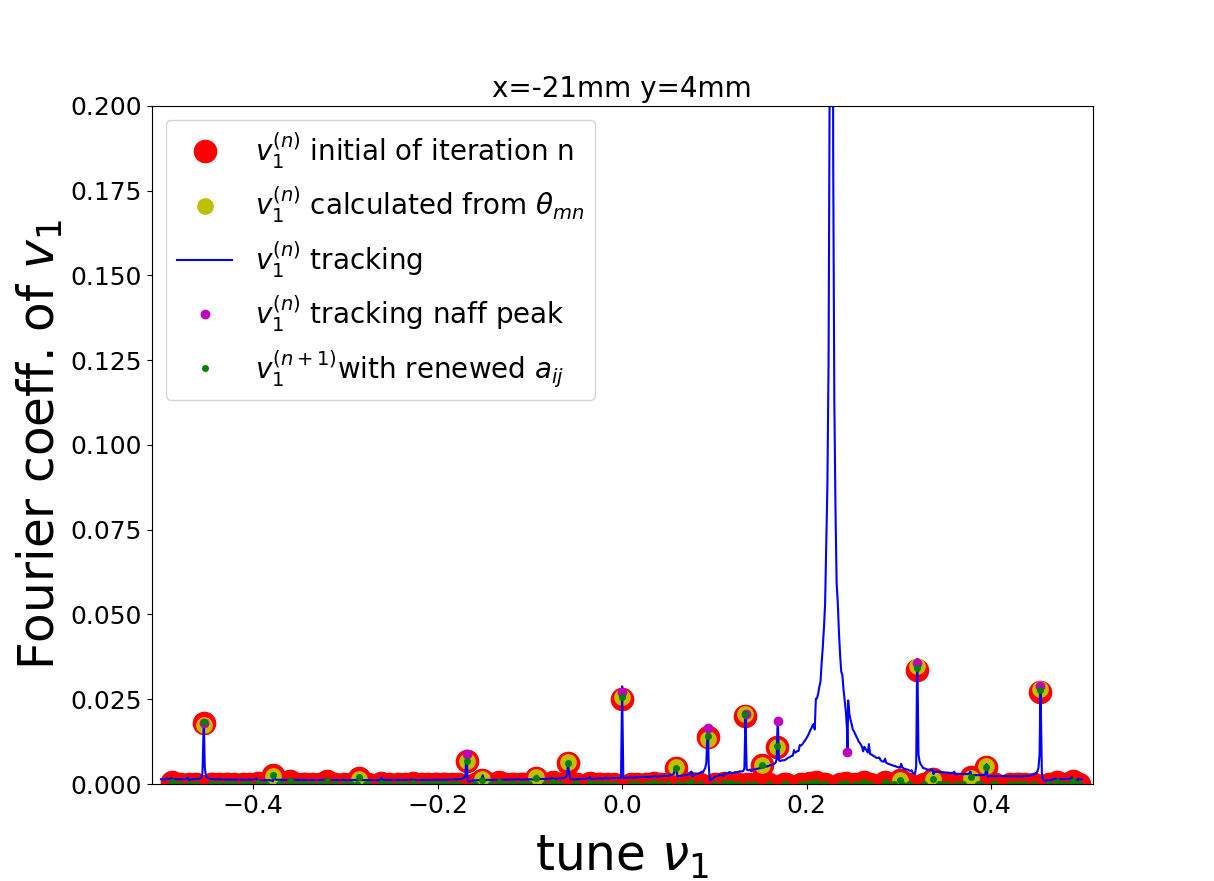}\includegraphics[width=0.5\columnwidth]{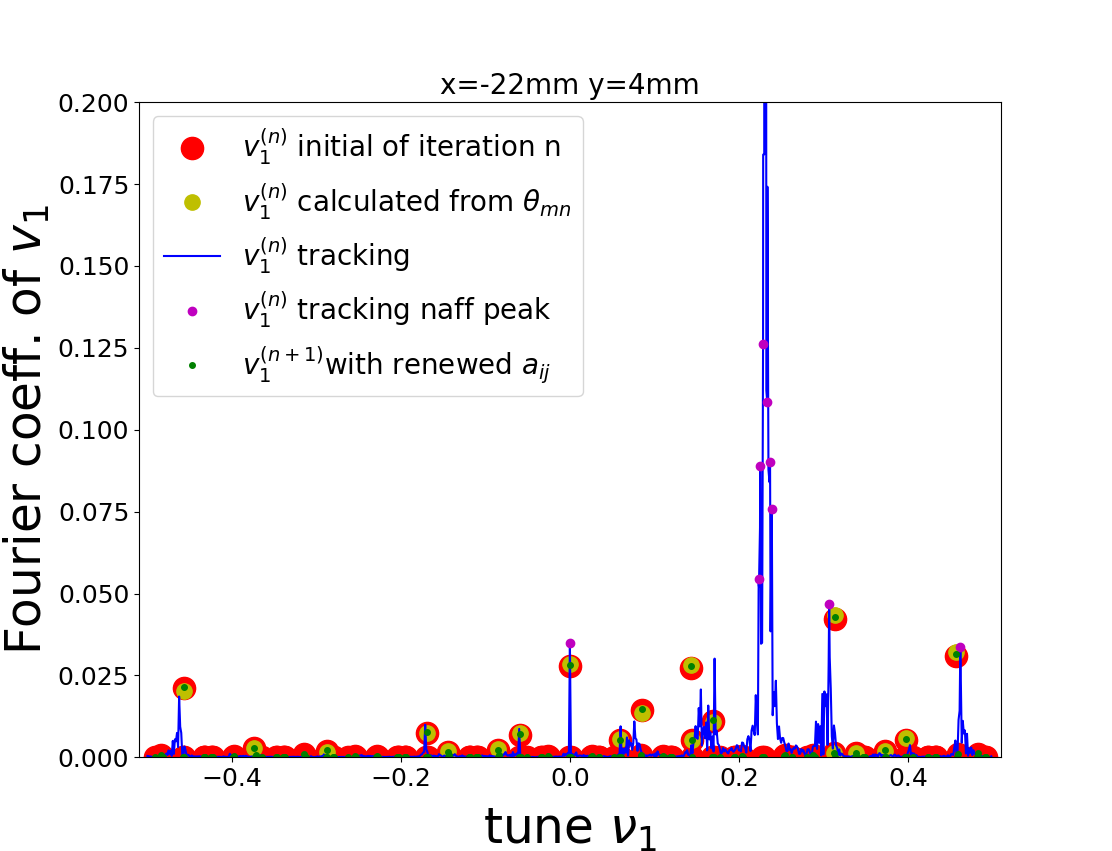}

\includegraphics[width=0.5\columnwidth]{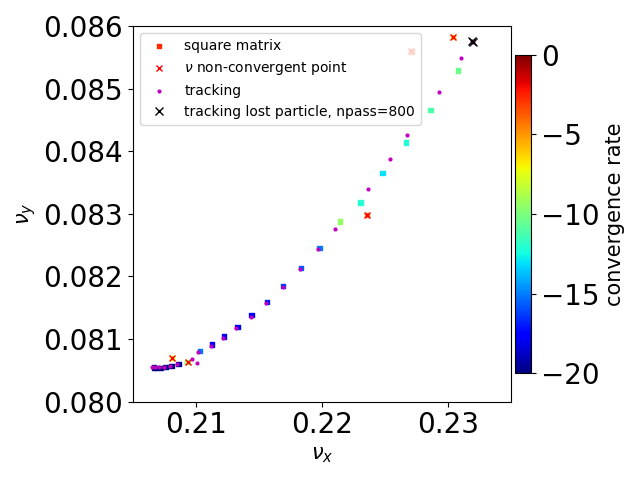}\includegraphics[width=0.5\columnwidth]{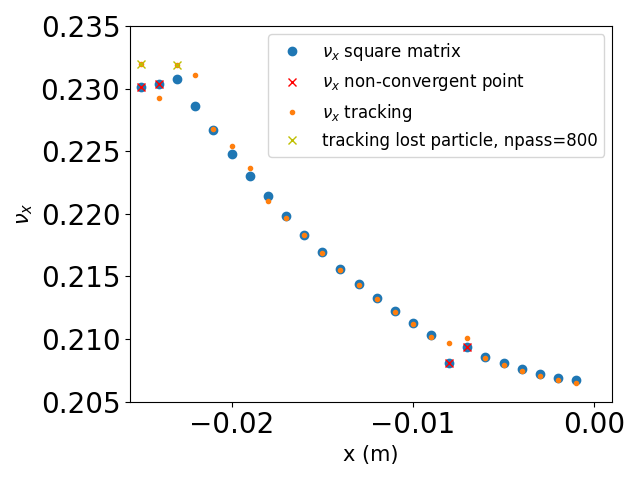}

\caption{(a) and (b): spectrum of $v_{1}^{(n)}$ at the last iteration $4$
(red, yellow, green) compared with tracking (FFT from tracking blue,
NAFF from tracking magenta) for $x=-21mm$ and $-22mm$ respectively.
In order to see the precision of the spectrum, the vertical limit
is 0.2 so the peak at nearly amplitude 1 is outside the scale and invisible.
(c) tune footprints in $\nu_{x},\nu_{y}$ plane. (d) $\nu_{x}$ vs.
$x$}

\end{figure}

In the following example of numerical application of the iteration
steps in Section \ref{sec:Perturbation}, we use the matrix $M$
derived from one of the lattices for NSLSII storage ring. To construct
the square matrix $M$, we first transform $x,p_{x},y,p_{y}$ to Courant-Snyder
variables $z_{x}=\overline{x}-i\overline{p}_{x}$, $z_{y}=\overline{y}-i\overline{p}_{y}$,
which are used to construct the monomial column $Z$. The the square
matrix $M$ is construct from lattice input file using Truncated Power
Series Algebra (TPSA) \cite{berz,dragt,forest1,forest,chao,bazzani}.
We use four polynomials $w_{x0},w_{x1},w_{y0},w_{y1}$ derived from
Jordan vector of power order 3, and corresponding linear combination
coefficients $\{a_{1j}^{(k)},a_{2j}^{(k)}\},j=1,2,3,4$. Initially
we take $a_{1}=1,0,0,0$, $a_{2}=0,0,1,0$. So initially we only use
$w_{x0},w_{y0}$ for $v_{1}$ and $v_{2}$ respectively. During the
iteration the contribution from $a_{12,}a_{13,}a_{14,}a_{21,}a_{22},a_{24}$
(starting from iteration $0$) increases to minimize the deviation of $v_{1}$
and $v_{2}$ from pure rotation, and improves the precision of $X^{(k)}$
so it is closer to real trajectory.

For a trajectory starting from $x=-22mm,y=4mm$, and momentum deviation $\delta =-0.025$
very close to dynamic aperture, when we take $n_{\theta}=12$, the
iteration leads to convergence as shown in Fig.6a. When we use tracking
by the code ELEGANT \cite{borland}
to calculate the one turn map from $X_{k}$ to $X_{k+1}$, as mentioned
in the definition of one turn mapping function $\phi_{1},\phi_{2}$
in Eq. \eqref{eq:vdef_phi}, the minimum of $\ln(\delta X_{n})$ (blue) is
-17 at the iteration 15 because the digital
noise (order of $e^{-17}=4\times10^{-8}$ mm) limited by the 7 digits
in ELEGANT output ascii file we used. When we detect the minimum we stopped the iteration at step 18.
Another way is to use square matrix of power order 5, the result is
the orange dots (``tpsa'') with the minimum of $\ln(\delta X_{n})$
at -29.4 at iteration 32 (order of $e^{-29.4}=1.7\times10^{-13}$
mm). Notice that even though the Jordan vector for the action-angle
variables is of power order 3, the one turn map can be exact, as given
by ELEGANT tracking. The results have
almost same convergence rate, and at the the iteration 18, the trajectory
difference is very small (order of $e^{-17}=4\times10^{-8}$ mm).
In the following, we we use square matrix of power order 3 to obtain
the 3rd order polynomials of Jordan form. However in the iteration
steps, the one turn map is calculated by the power order 5 square
matrix (the calculation is approximate) to study the iteration convergence
rate. As we mentioned before, the difference between using ELEGANT
(the precise method) or power order 5 square matrix (the approximate
method) is negligible in the optimization of the lattice.

For a scan from $x=$ -1mm to -26mm for every mm, we plot the minimum
$\ln(\delta X_{n})$ of iteration for each $x$ in Fig.6b, here $n_{\theta}=12$.
Because in Fig.6b our goal is only to study convergence, not to reach
very high precision for the orbit, the maximum iteration is set at
4. The blue curve at the top is the number of iterations reached vs.
$x$. For $x<-25mm$, and for $x=7,8$ mm the iteration diverges while
for other $x$ the iteration converges. The vertical red line and
light blue line in Fig.6b provide information about dynamical aperture
and the relation between divergence and $n_{\theta}$ , to be addressed
in Section V. The divergence at $x=7,8$ mm is due to resonance, where
the trajectories move around two 1D-tori and form two islands in the
4D phase space. The 1D-tori can also be calculated to very high precision
by square matrix method while trajectories in the resonance region
are organized around the 1D-tori. However, the study around resonance
region will be discussed in a separate publication. 

The same iteration for initial value over the $xy$ plane with initial
value of $p_{x},p_{y}=0,$ and momentum $\Delta p=0$ is shown in
Fig.1a (before the Introduction) with $n_{\theta}=12$. The minimum
of $\ln\delta X_{n}$ is represented by the color scale. The
white area represents divergence of the iteration. We refer this map
as a convergence map. A comparison of speed of the convergence map
calculation with that of the frequency map will be discussed in Section
VI. These two maps are entirely different maps, but both provide similar
space structure. This suggests that the convergence map can be used
in the nonlinear lattice optimization.

We use one point as an example of the iteration result, i.e. initial
value $x=-22mm,y=4mm,\delta=-2.5\%$ Fig.7a,b compare trajectory
calculated from ELEGANT and from iteration
4. Fig.7c,d show the trajectory in $v_{1}$ and $v_{2}$ complex plane
respectively. Dark blue dots represent initial trial $v_{1}^{(0)}$
,$v_{2}^{(0)}$ calculated with initial $\theta_{1}^{(0)}(\alpha),\theta_{2}^{(0)}(\alpha)$
and $n_{\theta}=12$. The red dots represent the result of $v_{1}^{(n)},v_{2}^{(n)}$
at the end of the iteration. The light blue lines represent $v_{1}^{(n)}$
,$v_{2}^{(n)}$ calculated from tracking trajectory $x,p_{x},y,p_{y}$
for 1024 turns. There is a very good agreement between tracking and
iteration results. 

The spectrum of $v_{1}^{(n)}$ for $x=-21mm$ and $-22mm$ are shown
in Fig.8 a,b. The main peaks are normalized to 1. The fluctuation peaks
(red $v_{1}^{(k)}$ , yellow calculated from $\widetilde{\theta}_{1hm}$
and green $v_{1}^{(n+1)}$ dots) agree with tracking (blue lines) with
peak (magenta dots) calculated by naff from tracking agree well even
with limited $n_{\theta}=12$ for $x=-21mm$. For $x=-22mm,$ the
difference is larger but the agreement is still very good considering
the particle lost at $N=5448$. In Fig 8.c,d the tune footprint in
$\nu_{x},\nu_{y}$ plane, and the $\nu_{x}$ vs. $x$ plot, the square
matrix tune at the last iteration agree well with tracking except
for points where the iteration diverges or particle lost in the tracking
(represented by crosses). 

In tracking for much longer time, the particle lost at $N=$5448 turns.
Similarly when $n_{\theta}$ increased to 17, the iteration diverges. 

There is a qualitative relation between the $n_{\theta}^{2}$ when
iteration diverges and the number of turns $N$ when particle lost,
obtained from numerical experiences. We have some intuitive understanding
of this relation, but lack an analytical analysis so far, as will
be addressed later in Section VII.

\section{\label{sec:Optimization}An Example of Nonlinear Lattice Optimization }

As a practical example for the utility of convergence maps (CMs),
we used optimization of harmonic sextupoles for NSLS-II to maximize
its on-momentum dynamic aperture (DA). 

The lattice used for this was one super-period of NSLS-II (15 super-periods
in the whole ring) without any insertion devices (often referred to
as `` bare lattice''). The knobs for
this optimization problem were the strengths for all 6 families of
harmonic sextupoles. For each set of sextupole values, the DA, defined
to be the maximum radius within which the convergence value stays
below -12, was searched for each radial line. Nine radial lines covered
the upper half-plane of x-y initial coordinate space. Each radial
line search progressed monotonically outward with a step size of 1
mm, and stopped once the threshold convergence value was exceeded.
These radial DA values were then used directly as the multi-objectives
for the optimization problem. 

The optimization algorithm employed for this problem was MOGA (multi-objective
genetic algorithm) \cite{MOGA, yang} implemented
with DEAP Python package \cite{fortin, DEAP2}. 

Figure 9a shows the frequency map (FM) of one of the optimal lattices
after CM optimization. As in the previous section, the FMs in this
section were generated by the ``frequency map''
command of ELEGANT \cite{borland}.
The magenta circles correspond to the 9 radial apertures found by
CM during the optimization process. The full convergence map for the
same lattice is shown in Fig. 9b, whose boundary looks similar to
that of the FM. The horizontal aperture (near y=0) extends up to -30
mm and +35 mm, while satisfying the minimum required 2-mm vertical
DA. In this sense, this optimized lattice appears to be better than
the  NSLS-II bare lattice whose FM and CM are shown in Fig. 1a,1b. 

\begin{figure}[!htbp]
\includegraphics[width=0.5\columnwidth]{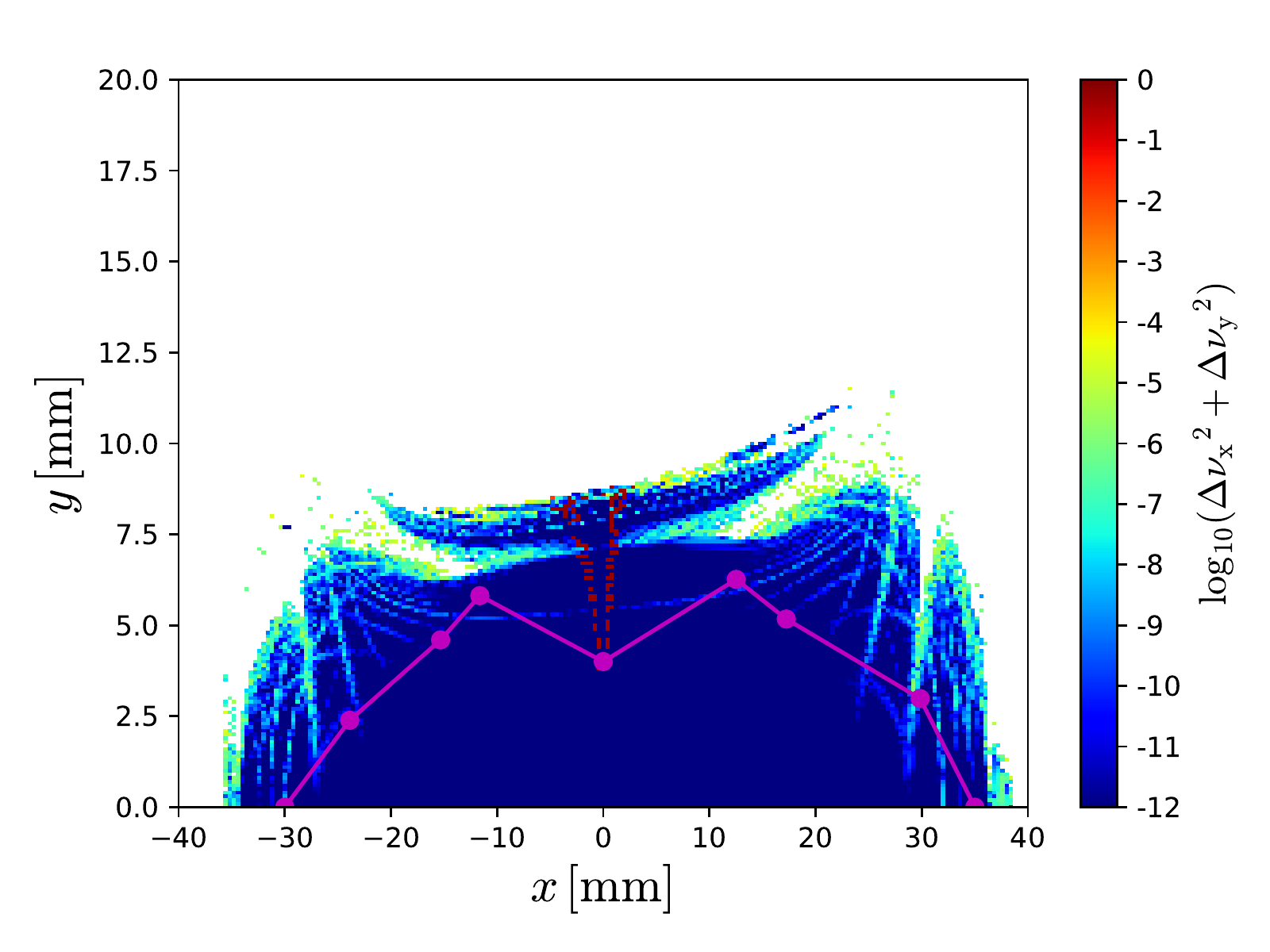}\includegraphics[width=0.5\columnwidth]{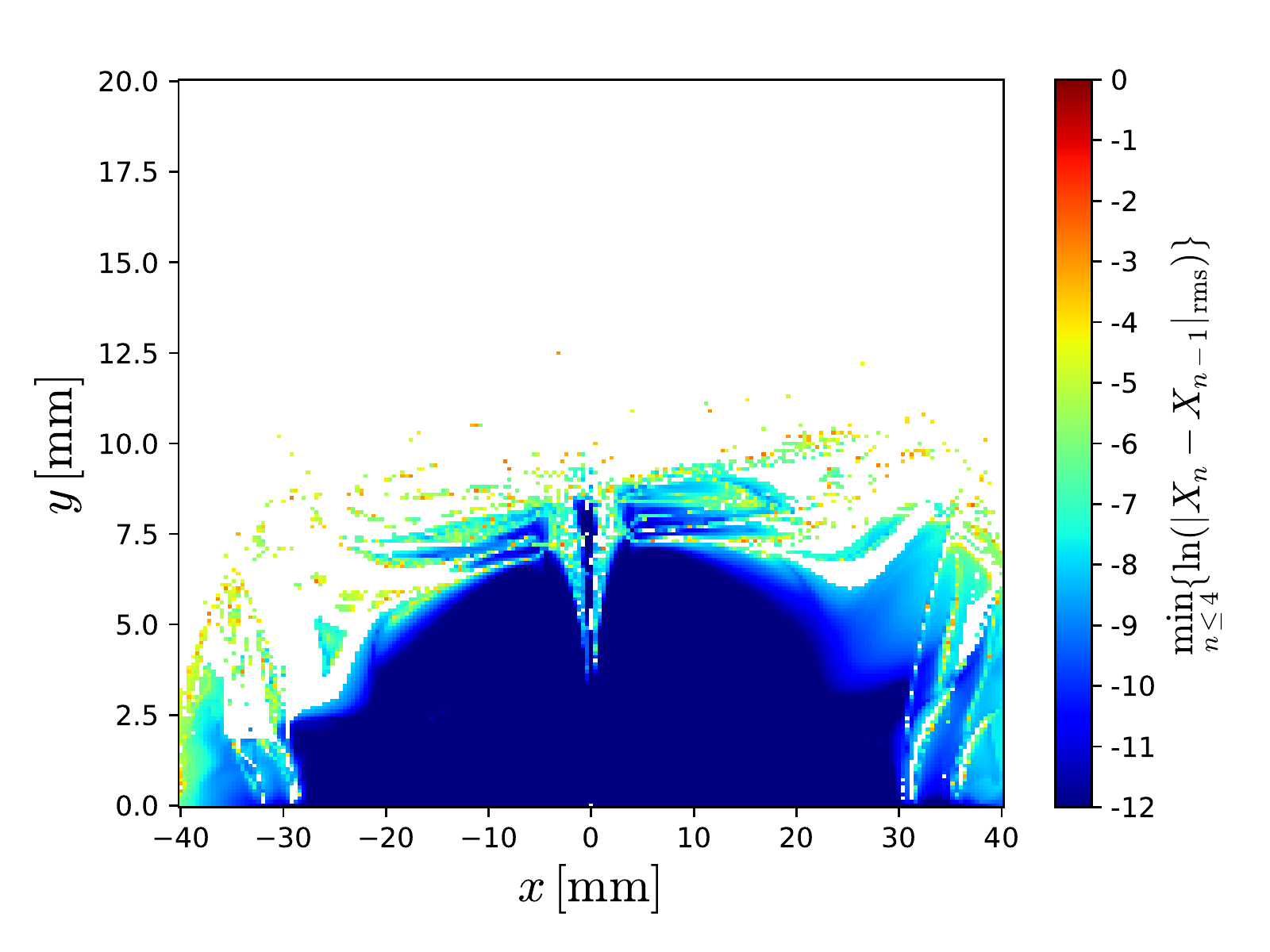}

\caption{(a)The frequency map and the radial apertures (magenta circles)
found by convergence map for one of the optimized lattices. (b) The
full convergence map for the same lattice and on the same grid used
in (a).}

\end{figure}

This CM optimization was also able to find a lattice, shown in Fig.10
whose FM and CM appear very similar to those of NSLS-II bare lattice
shown in Fig. 1. These two optimized lattices shown in Fig.9 and Fig.10
demonstrate that the optimization based on CM can find lattices at
least as good as or better than the optimization using DA based on
many-turn particle survival. Furthermore, it achieves this feat with
only a fraction of the computation resources.

\begin{figure}[!htbp]
\includegraphics[width=0.5\columnwidth]{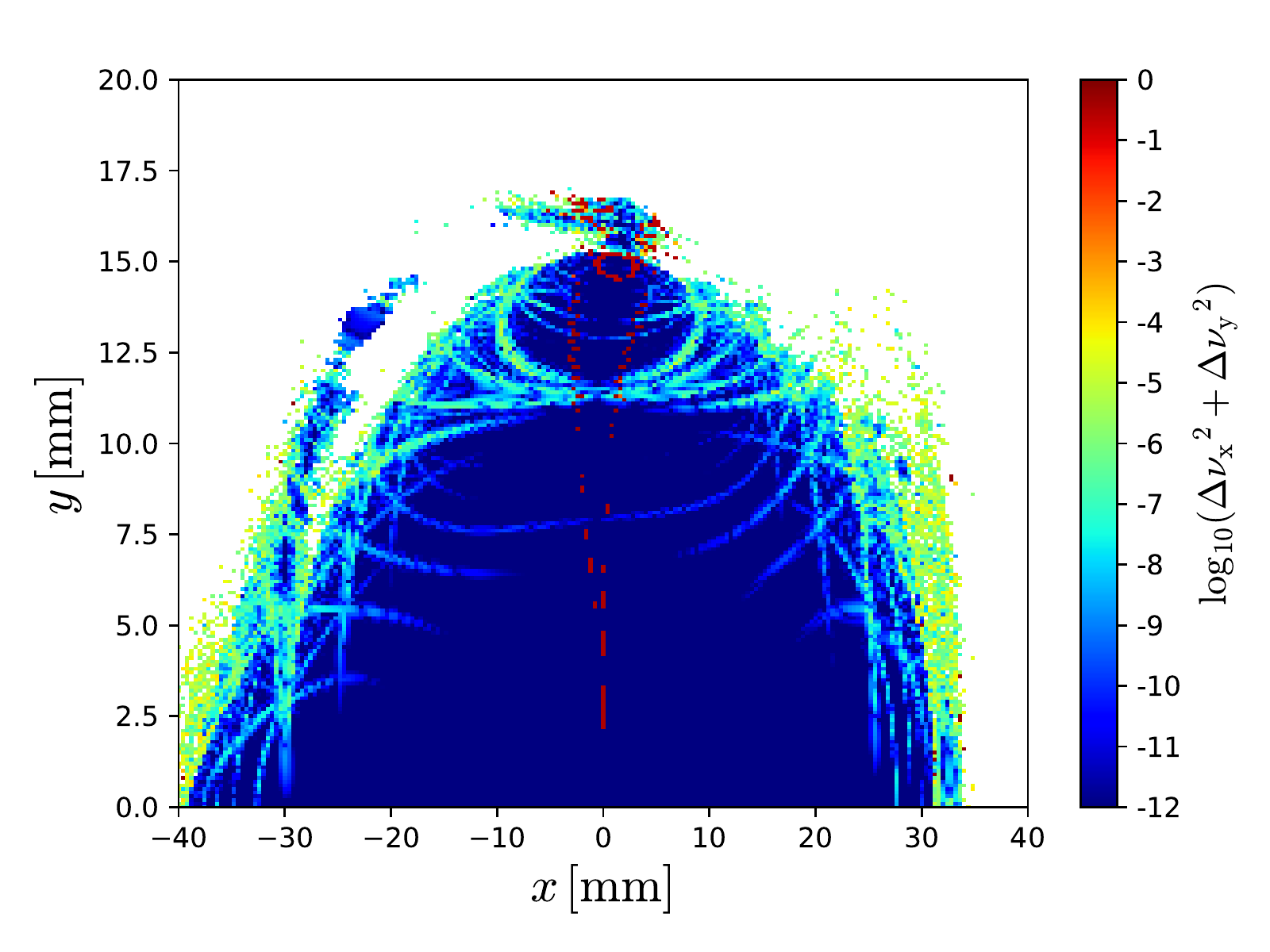}\includegraphics[width=0.5\columnwidth]{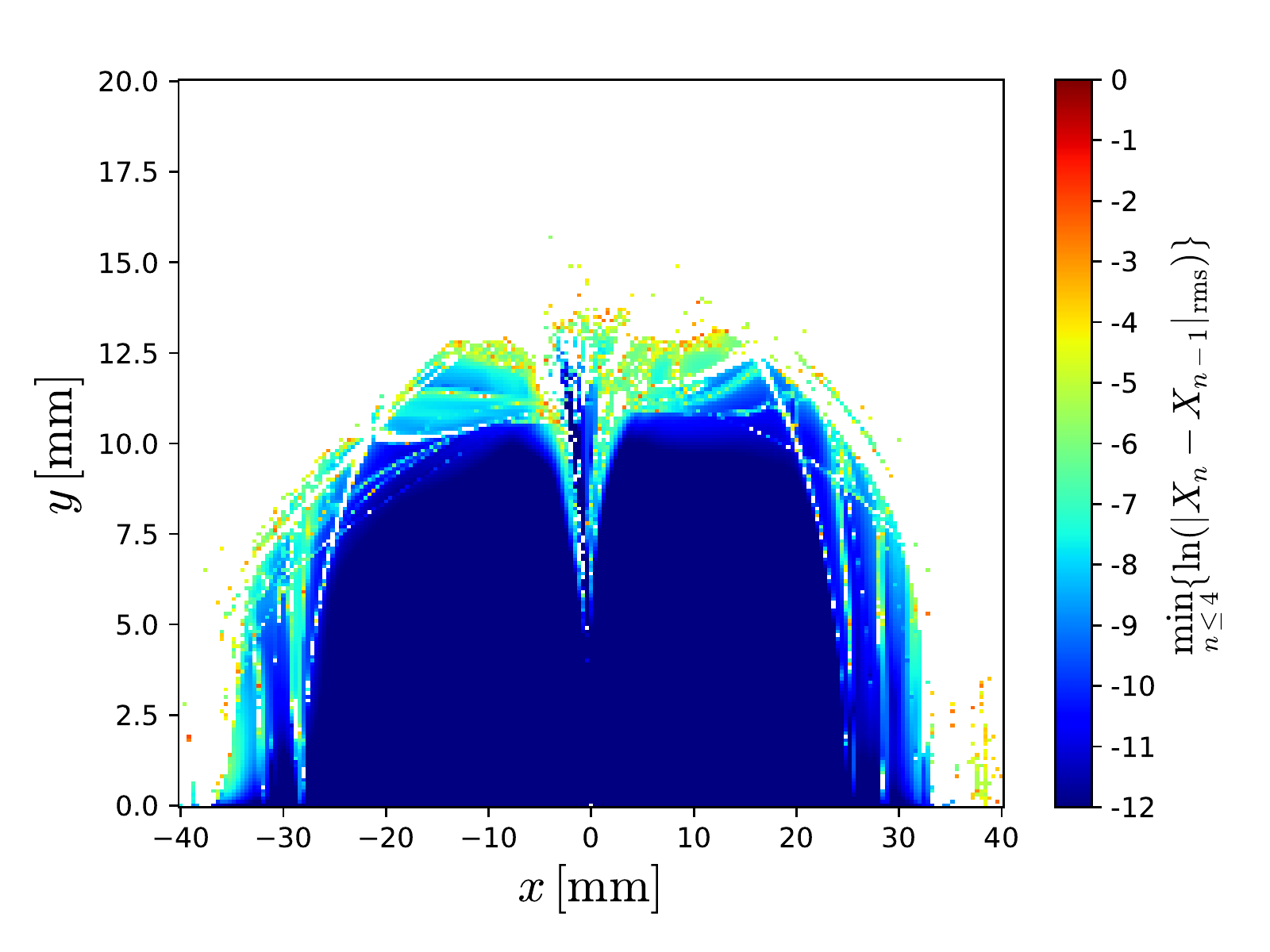}

\caption{(a)The frequency map and (b) the convergence map for  an optimized lattice similar to NSLS-II bare lattice}

\end{figure}

\section{\label{sec:ComputationTime}Computation Time Comparison of Convergence
Map vs. Frequency Map}

Computation time was compared between the convergence map and the
frequency map using one super-period (2 cells) and the whole ring
(15 super-period) of the NSLS-II bare lattice. By ``bare'' ,
it means there is no insertion device element in the lattice. 

To compare the two maps, we need to compare the computation time for
selected points in x-y plane in the NSLS-II bare lattice. If we choose
the points in an unstable region, some particles may be lost during
tracking. This would make the comparison difficult. 

For a fair comparison, we chose a stable region. For both types of
maps, an initial coordinate region of $+10 \le x[\text{mm}]
\le +11$ and $+1 \le y[\text{mm}]
\le +2$ was selected as particles launched from this region
are very stable and can last at least 1024 turns specified for frequency
map analysis. This square region was divided into $2\times2$, $3\times3$,
$5\times5,10\times10,50\times50,100\times100$ grid points as a set
of different number of points. Each grid point is used as an initial
transverse coordinate for both maps. The momentum offset was zero. 

For frequency map computations, we used ELEGANT
``frequency\_map'' command \cite{borland}
to compute the diffusion defined by the tune changes between the first
512 and the latter 512 turns. 

For convergence map computations, PyTPSA \cite{PyTPSA}
was used to create truncated power series (TPS) objects and handle
all the algebraic operations on them while the TPS objects are propagated
through all the lattice elements in a Python module where the symplectic
integration method of TRACY \cite{tracy} has been reimplemented. 
The components of the TPS have been confirmed with simulation code MADX-PTC \cite{skowronski}.
During the speed test we use $n_{\theta}=12$. The number of iteration
is set at 4. The polynomials $w_{j}=u_{j}Z,$ based on Jordan form
as introduced by Eq. \eqref{eq:w_transformation}, are polynomials of 3rd power
order. For readers who might be interested in the detailed implementation
of our method, please see \cite{sqmxcode}.

All the computations in this section were performed using a single
core of Intel Xeon Gold 6252 CPU at 2.10 GHz (hyper-threading enabled).
The results are shown in Fig. 2 in the Introduction.This proposed
CM method is also friendly to parallelization, which has been demonstrated
to scale well to 128 cores.

The computation time of frequency maps (FM) linearly scaled with the
number of grid points as expected. It was also expected to linearly
scale with the number of super-periods (SP), as each point requires
tracking of a single particle from the beginning to the end of the
selected lattice. Thus, the whole-ring lattice should have taken roughly
15 times longer than the 1-SP lattice. However, the time only increased
by 10.5. This appears to indicate the overhead of non-tracking portion
of ELEGANT code is not negligible, compared
to the tracking portion. 

The most notable feature of the convergence map (CM) time is the fact
that it changed very little for the case of $10^{4}$ points whether
the lattice was 1 or 15 super-periods. This makes sense because once
the TPSA calculation for a lattice is finished at the beginning, the
computation cost is the same for each grid point, whether the lattice
was 1 or 15 super-periods, unlike the tracking-based FM whose computation
time is proportional to the length of the lattice. Note that the initial
TPSA calculation does depend on the length of the lattice. However,
it only increased from 0.81 s for 1 SP to 6.45 s for 15 SP. In both
cases, this initial setup time is tiny compared to the total time
of 100 seconds it took to compute the convergence values for $10^{4}$
points.

The speed of CM for $10^{4}$ points was 31.2 times faster than that
of FM for the 1-SP case, while it was 314 times faster for the 15-SP
case. These speed improvement factors include all the overhead and
initial setup times. However, the advantage of CM diminishes as the
number of points decrease, since the initial TPSA computation time
starts to dominate the total CM computation time. Therefore, CM is
particularly useful when the number of initial coordinate points whose
stability needs to be investigated is quite large and/or when the
lattice under study is very long and complex (e.g., lattices with
multipole and alignment errors included and lattices with no periodicity
such as colliders). 

The dashed lines in Fig. 2 shows the computation times for CM without
the initial setup times. Both the 1-SP and 15-SP curves show good
linearity with the number of grid points. They are also almost on
top of each other. This demonstrates the earlier statement of the
convergence value computation time being independent of the lattice
length/complexity, as long as the initial TPSA computation time is
excluded. 

\section{\label{sec:DynamicAperture}Survival turn number $N$ and Dynamic
Aperture, and its Relation to Convergence-Divergence-$n_{\theta}$
Dependence }

Frequency map and convergence map are very different but related.
In a frequency diagram, the dynamical aperture is given by the boundary
where the particle is lost within a specified number of turns $N$.
To find dynamical aperture defined by the divergence of the iteration
by square matrix method, we need to understand the relation between
divergence of the iteration and $n_{\theta}$. 

The survival turn number is very sensitive to initial position $x$,
so the study is based on statistical average. In Fig.11.a, the number
of survival turns is plotted vs. $x$ for every 0.1mm, and for every
0.1mm with neighbour 20 points separated by $10\mu m$, tracking $N=65536$
turns. There is a boundary at $x=$-21.2mm if we choose $N=60000$,
and the very thin area at $x=-22.8mm$. But for application in light
source, with damping time about 10ms, if we take $N=1024$, then Fig.11.b
(the same plot as Fig.11a with vertical range reduced to 10000 ) shows
fluctuation of the number of survival turns is so large, that we need
to further average over a certain range of $x$. Fig.11c shows the
result of averaging over every 20 points of neighbour $x$ and compared
with the same set of data in Fig.11b, the dynamical aperture is about
$x=-23mm$. Similar plot is shown in Fig.11d with more points of average
gives less fluctuation.

\begin{figure}[!htbp]
\includegraphics[width=0.5\columnwidth]{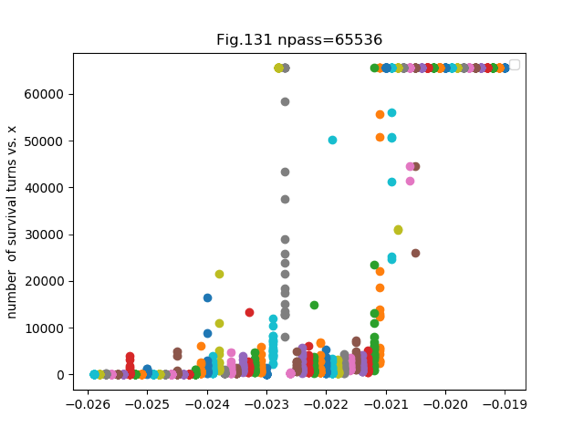}\includegraphics[width=0.5\columnwidth]{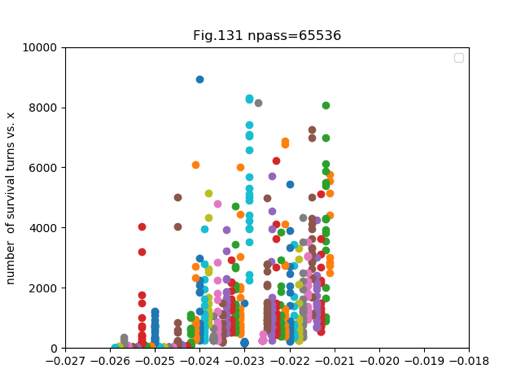}
\includegraphics[width=0.5\columnwidth]{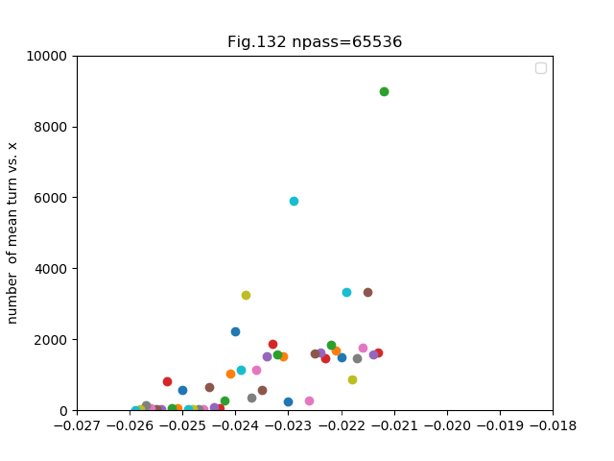}\includegraphics[width=0.5\columnwidth]{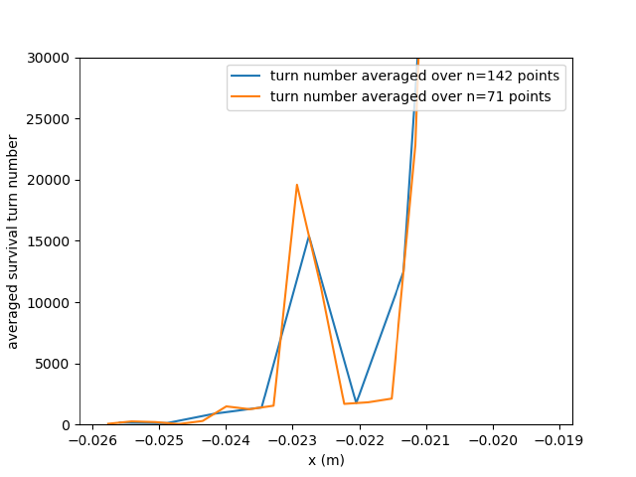}

\caption{a and b: number of survival turns vs. $x$ by tracking ELEGANT
in 70000 and 10000 scale c: same plot averaged over every 20 points
of neighbour $x$ d: same plot averaged over 0.35mm (orange) and 0.7
mm (blue) respectively}
\vspace{-1.0em}
\end{figure}

Similarly, we can use the convergence-divergence-$n_{\theta}$ dependence
in iteration by square matrix method to estimate the dynamical aperture.
Fig.12 a,b are the 3D plot of iteration convergence rate vs. $n_{\theta}$
for $x=-20.1mm,20.2mm$ respectively. At $x=-20.1mm$ the iterations
are convergent from $n_{\theta}=25$ to $80$ with only exception
at $n_{\theta}=68,72$. But at $x=-20.2mm$ the iterations diverge
for all $n_{\theta}>35$. Numerical study for many different $x$
shows when $n_{\theta}$ increases above a certain number, the divergence
points form a continuous band with only very few points convergent. 

The lowest point $n_{\theta}$ of the divergence band is also sensitive
to $x$. Similar to tracking, Fig.12c plot the points of $n_{\theta}^{2}$
where the iteration converges. The distribution is also sensitive
to the initial $x$, and Fig.12b shows at $x=-20.2$ there are band
of divergence points above $n_{\theta}>40$. To be able to estimate
dynamical aperture from this data, again, the $n_{\theta}^{2}$ is
averaged over a small range of $x$ for every point, and plotted in
Fig.12d. Compared with Fig.11d, Fig.12d also show that there is fast
decrease of convergence at $x=-20.2mm$. If we take divergence at
$n_{\theta}>25$ for aperture, then the aperture is estimated at $x=-21mm$.
For crude estimate we may take $n_{\theta}=12$, then the aperture
would be estimate as $x=-22mm$. This example indicates that even
with relatively low $n_{\theta}=12$ the dynamical aperture is within
$1mm$ from the aperture obtained by tracking of 6000 turns. 

There is a resonance line at $x=-18.6mm$, an indication the iteration
convergence is sensitive to resonance. A more detailed calculation
leads to convergence at this point, but the calculation takes some
more time. Since the main goal of this paper is to study dynamical
aperture, the resonance study will be addressed in future publication.
As we mentioned in regard of Fig.6b, near the resonance center, the
square matrix method can also be applied to obtain very accurate
information about a 1D-torus. Even though qualitatively we can use
the convergence map to estimate the dynamical aperture, we still lack
a more quantitative analytical understanding of the relation between
iteration convergence at $n_{\theta}$ and particle loss at turn $N$.
The fact that Fig.12c seems to be a little more regular than Fig.11b
indicates that there might be some analytical way to explain the statistics
of divergence vs. $n_{\theta}.$

\begin{figure}[!htbp]
\includegraphics[width=0.5\columnwidth]{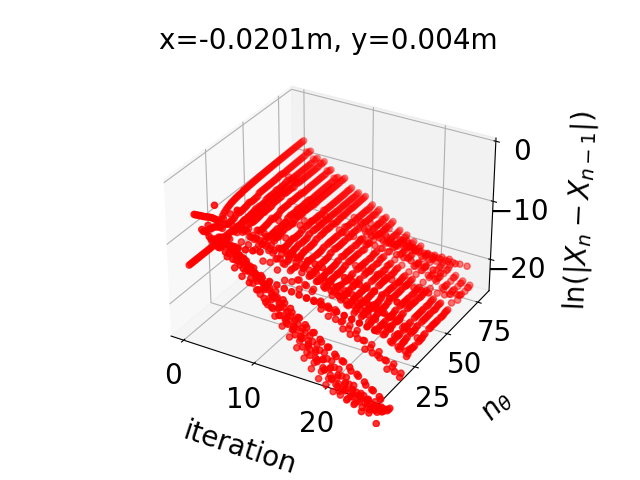}\includegraphics[width=0.5\columnwidth]{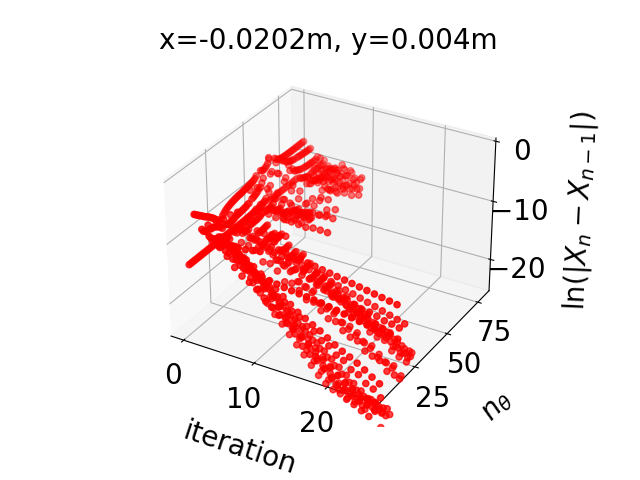}
\includegraphics[width=0.5\columnwidth]{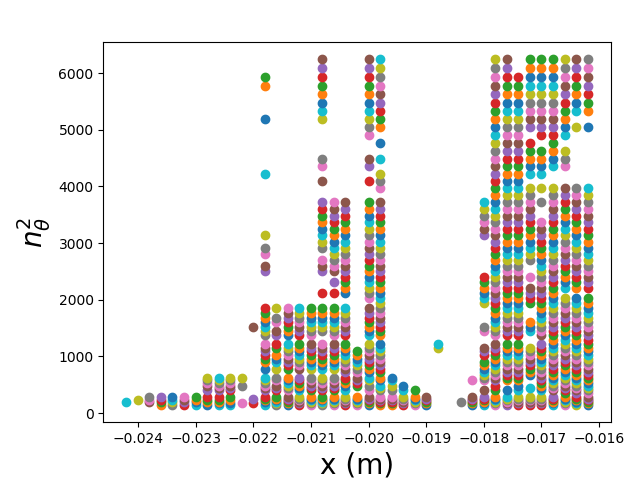}\includegraphics[width=0.5\columnwidth]{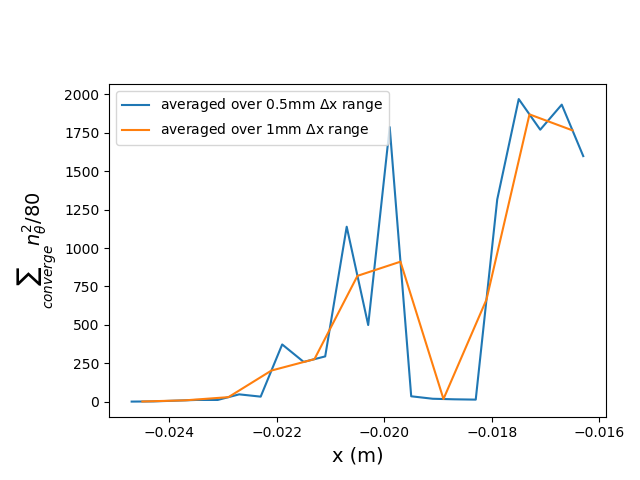}

\caption{(a) the 3D plot of iteration convergence rate vs. $n_{\theta}$ for
a: $x=-20.1mm.(b),x=-20.2mm$. (c) $n_{\theta}^{2}$ vs. x. (d) $n_{\theta}^{2}$
averaged over 80 points of neighbour $x$.}
\vspace{-1.0em}
\end{figure}

The comparison of Fig.12d and Fig.11d indicates the possibility of
using convergence map to study dynamical aperture. Hence the speed
of the calculation is important, as discussed in Section V.

\section{\label{sec:Conclusion}Conclusion}

In this paper we show that the action-angle variables derived from
square matrix method is close to a pure rotation, hence it is possible
to rewrite the nonlinear dynamical equations in terms of these variables
as an exact equation. The equation is in the form of pure rotation
with nonlinear terms as perturbation. Hence an iteration steps developed
using perturbation method to solve the nonlinear dynamical equation
are convergent up to dynamic aperture or the border of resonance region.
The convergence rate varies depending on how close the trajectory
is to the dynamic aperture or resonance region. Hence the convergence
rate is a function of the initial particle coordinates. For example
the convergence rate can be plotted as a function of horizontal
and vertical coordinates, as a color map. This ``convergence map''
can be used to study the stability of the nonlinear system.

This convergence map is similar but very different from frequency
map calculated by tracking. The results agree with tracking well on
dynamic aperture, tune footprint and phase space trajectory, and frequency
spectrum to high precision. Using an NSLS-II lattice as an example,
we carried out an extensive comparison of the optimization by traditional
tracking method with the convergence map. We compared the speed and
the quality of the optimization result, and show that depends on the
complexity of the lattices. The iteration method is about 30 to 300
times faster than tracking.

Hence the convergence map is suitable for nonlinear optimization of
storage ring lattice, and in particular for the study of the very
long term behavior in storage rings with very large number of sextupoles
or high order multipoles.

\section*{Appendix A Fourier Transform Solution of Iteration Equation }

We write the two dimensional Fourier transform of 
$\phi_{1}(\alpha_{1},\alpha_{2}),\phi_{2}(\alpha_{1},\alpha_{2}),
\theta_{1}(\alpha_{1},\alpha_{2}),\theta_{2}(\alpha_{1},\alpha_{2})$
in Eq. \eqref{eq:dynamics_theta_k} (the iteration number $k$ is implicitly implied) as:
\begin{equation}
\begin{split} & \phi_{1}(\alpha_{1},\alpha_{2})=\omega_{1}+\sum_{\tiny{\begin{matrix}m,n\\
|n|+|m|\neq0
\end{matrix}}}^{n_{\theta},n_{\theta}}\widetilde{\phi}_{1nm}e^{in\alpha_{1}}e^{im\alpha_{2}},\phi_{2}(\alpha_{1},\alpha_{2})=\omega_{2}+\sum_{\tiny{\begin{matrix}m,n\\
|n|+|m|\neq0
\end{matrix}}}^{n_{\theta},n_{\theta}}\widetilde{\phi}_{2nm}e^{in\alpha_{1}}e^{im\alpha_{2}}\\
 & \theta_{1}(\alpha_{1},\alpha_{2})=\alpha_{1}+\sum_{n,m=0,0}^{n_{\theta},n_{\theta}}\widetilde{\theta}_{1nm}e^{in\alpha_{1}}e^{im\alpha_{2}},\theta_{2}(\alpha_{1},\alpha_{2})=\alpha_{2}+\sum_{n,m=0,0}^{n_{\theta},n_{\theta}}\widetilde{\theta}_{2nm}e^{in\alpha_{1}}e^{im\alpha_{2}}
\end{split}
\label{eq29}
\end{equation}

where in the sum in $\phi_{1},\phi_{2}$ the indexes run from $0$
to $n_{\theta}$ except the term for $m=n=0$, i.e., the constant
terms are removed and replaced by $\omega_{1},\omega_{2}$ respectively.
Compare both sides of the Fourier transform of the first equation
in Eq. \eqref{eq:dynamics_theta} leads to

\begin{align*}
\mathop{\mathop{}} & \alpha_{1}+\omega_{1}+\sum_{n,m=0,0}^{n_{\theta},n_{\theta}}\widetilde{\theta}_{1nm}e^{in(\alpha_{1}+\omega_{1})}e^{im(\alpha_{2}+\omega_{2})}-\alpha_{1}-\sum_{n,m=0,0}^{n_{\theta},n_{\theta}}\widetilde{\theta}_{1nm}e^{in\alpha_{1}}e^{im\alpha_{2}}\approx\omega_{1}+\sum_{\tiny{\begin{matrix}m,n\\
|n|+|m|\neq0
\end{matrix}}}^{n_{\theta},n_{\theta}}\widetilde{\phi}_{1nm}e^{in\alpha_{1}}e^{im\alpha_{2}}
\end{align*}

\begin{align*}
 & \sum_{n,m=0,0}^{n_{\theta},n_{\theta}}\widetilde{\theta}_{1nm}e^{in(\alpha_{1}+\omega_{1})}e^{im(\alpha_{2}+\omega_{2})}-\sum_{n,m=0,0}^{n_{\theta},n_{\theta}}\widetilde{\theta}_{1nm}e^{in\alpha_{1}}e^{im\alpha_{2}}\approx\sum_{\tiny{\begin{matrix}m,n\\
|n|+|m|\neq0
\end{matrix}}}^{n_{\theta},n_{\theta}}\widetilde{\phi}_{1nm}e^{in\alpha_{1}}e^{im\alpha_{2}}
\end{align*}

\begin{align*}
 & \sum_{n,m=0,0}^{n_{\theta},n_{\theta}}\widetilde{\theta}_{1nm}e^{i(n\omega_{1}+m\omega_{2})}e^{in\alpha_{1}}e^{im\alpha_{2}}-\sum_{n,m=0,0}^{n_{\theta},n_{\theta}}\widetilde{\theta}_{1nm}e^{in\alpha_{1}}e^{im\alpha_{2}}\approx\sum_{\tiny{\begin{matrix}m,n\\
|n|+|m|\neq0
\end{matrix}}}^{n_{\theta},n_{\theta}}\widetilde{\phi}_{1nm}e^{in\alpha_{1}}e^{im\alpha_{2}}
\end{align*}

\begin{align*}
 & \widetilde{\theta}_{1nm}e^{i(n\omega_{1}+m\omega_{2})}-\widetilde{\theta}_{1nm}\approx\widetilde{\phi}_{1nm}
\end{align*}

\begin{align}
 & \widetilde{\theta}_{1nm}=\frac{\widetilde{\phi}_{1nm}}{e^{i(n\omega_{1}+m\omega_{2})}-1}\nonumber \\
 & \widetilde{\theta}_{2nm}=\frac{\widetilde{\phi}_{2nm}}{e^{i(n\omega_{1}+m\omega_{2})}-1}\label{eq:thetatildw}
\end{align}

for $\text{ (}|n|+|m|\neq0),$i.e, except the constant terms. For
the constant terms, both sides are zeros so there we still need to
find $\widetilde{\theta}_{100},\widetilde{\theta}_{200}$ from other
condition. They are determined by the condition that the second line
of Eq. \ref{eq29}  must be valid for $\alpha_{1}\equiv k\omega_{1}+\theta_{10},\alpha_{2}\equiv k\omega_{2}+\theta_{20}$
with all $k$ including $k=0$, i.e., when $\theta_{10}(\theta_{10},\theta_{20})=\theta_{10},\theta_{20}(\theta_{10},\theta_{20})=\theta_{20}$
so

\begin{equation}
\begin{split} & \theta_{10}=\theta_{10}+\sum_{n,m=0,0}^{n_{\theta},n_{\theta}}\widetilde{\theta}_{1nm}e^{in\theta_{10}}e^{im\theta_{20}}\text{ }\\
 & \theta_{20}=\theta_{20}+\sum_{n,m=0,0}^{n_{\theta},n_{\theta}}\widetilde{\theta}_{2nm}e^{in\theta_{10}}e^{im\theta_{20}}.
\end{split}
\label{theta expansion-1-1}
\end{equation}
Hence

\begin{equation}
\begin{split} & \widetilde{\theta}_{100}=-\sum_{\tiny{\begin{matrix}m,n\\
|n|+|m|\neq0
\end{matrix}}}^{n_{\theta},n_{\theta}}\widetilde{\theta}_{1nm}e^{in\theta_{10}}e^{im\theta_{20}}\\
 & \widetilde{\theta}_{200}=-\sum_{\tiny{\begin{matrix}m,n\\
|n|+|m|\neq0
\end{matrix}}}^{n_{\theta},n_{\theta}}\widetilde{\theta}_{2nm}e^{in\theta_{10}}e^{im\theta_{20}}
\end{split}
\label{theta expansion-1-1-1}
\end{equation}

During the $k^\text{th}$ iteration, $\widetilde{\phi}_{1nm},\widetilde{\phi}_{1nm},\omega_{1},\omega_{2}$
of the right hand side should be label as $(k)$, while $\widetilde{\theta}_{1nm},\widetilde{\theta}_{2nm}$
of the left hand side should be labeled as $(k+1)$, as labeled in Eq. \eqref{eq:dynamics_theta_k}.
Thus the inverse Fourier transform gives the solution $\theta_{1}^{(k+1)}(\alpha_{1},\alpha_{2}),\theta_{2}^{(k+1)}(\alpha_{1},\alpha_{2})$.
In the numerical calculation, $\alpha_{1},\alpha_{2}$ are only specified
at discrete $n_{\theta}\times n_{\theta}$ points on the $\alpha_{1},\alpha_{2}$
torus plane of period $(2\pi\times2\pi)$. 

\section*{\label{sec:CalLin}Appendix B Calculation of Linear Combinations
using a known trajectory}

Near the elliptical fixed point, the dynamics is dominated by the
linear terms $z_{x},z_{y}$ in the square matrix $M,$ so $w_{x0}(X)$
and $w_{y0}(X)$ are near exact action-angle variables, and carry
out nearly a pure rotation independently with linear tune $\mu_{x},\mu_{y}$
respectively. The coefficients $\{a_{ij}\}$ in Eq.(\ref{eq:vdef_linear})
can be chosen as $a_{11}^{(0)}=1,a_{23}^{(0)}=1$ while all other
$a_{ij}^{(0)}=0$ for the first iteration step. In the case of increased
amplitude, these coefficients are determined by $x$,$p_{x}$,$y$,$p_{y}$
in a larger neighborhood near the fixed point. Obviously, the high
power terms in Eq.(\ref{eq:uz_umz_0}) serve as a perturbation to the
rigid rotation. (For a much more detailed and rigrorous description
we refer to Poincare-Birkhoff theorem \cite{brown}). During the
iteration, the higher power terms in $w_{x1}(X),w_{y0}(X),w_{y1}(X)$
contribute to the deviation of $v_{1}$ from a pure rotation, and the
same way $w_{x0}(X),w_{x1}(X),w_{y1}(X)$ contribute to the deviation
of $v_{2}$. Hence $a_{ij}^{(0)}$ can be further minimized by a least
square method to $a_{ij}^{(1)}$ as a starting point of the second
iteration step. This can be continued for every iteration step $k$
for $a_{ij}^{(k)}$. Our experience shows renew $a_{ij}^{(k)}$ in
each iteration step makes the convergence faster.

In the following we shall show that if we have a numerical direct
integration of the dynamical equations, i.e., if we have the trajectory
$X^{(k-1)}$, we can use the Fourier expansion of $w_{xi}(X^{(k-1)}),w_{yi}(X^{(k-1)})$
to determine the linear combinations $a_{ij}^{(k)}$ that minimize
the deviation from pure rotation for the approximate rigid rotation
$v_{1},v_{2}$. Hence an approximate trajectory can be used to determine
the linear combinations by a least square method.

For a trajectory $\theta_{1}^{(k-1)}(\alpha_{1},\alpha_{2}),\theta_{2}^{(k-1)}(\alpha_{1},\alpha_{2})$,
i.e., the solution of Eq. \eqref{eq:dynamics_theta_k}, the coordinates
$X^{(k-1)}$ can be found by the inverse function of Eq. \eqref{eq:vdef_linear},
as functions of $\alpha_{1},\alpha_{2}$ (modulo $2\pi$), hence the
eigenvectors $w_{x0}$ , $w_{x1}$, $\ldots$ can also be Fourier
expanded in terms of $\alpha_{1},\alpha_{2}$. 

For simplicity in writing, if we choose $n_{v}$ eigenvectors for
the linear combinations, we label them as $w_{j}$ with $j=1,2,..,n_{v}$.
For example, for Eq.(\ref{eq:vdef_linear}), $n_{v}=4$, $w_{1}\equiv w_{x0},w_{2}\equiv w_{x1},$
$w_{3}\equiv w_{y0},w_{4}\equiv w_{y1}$. We have the expansion

\begin{equation}
w_{j}(\alpha_{1},\alpha_{2})=\sum_{n,m}\widetilde{w}_{jnm}e^{in\alpha_{1}}e^{im\alpha_{2}}\qquad(j=1,2,..n_{v})\label{eq:10}
\end{equation}
Here $m,n=0,1,2,...,n_{\theta}$ see Eq.(\ref{eq29}). Now we look
for linear combinations $a_{1j}$,$a_{2j}$ to construct the two approximate
action-angle variables $v_{1},v_{2}$
\begin{align}
v_{l} & =\sum_{j=1}^{n_{v}}a_{lj}w_{j}=\sum_{n,m}\left(\sum_{j=1}^{n_{v}}a_{lj}\widetilde{w}_{jnm}\right)e^{in\alpha_{1}}e^{im\alpha_{2}}\label{eq:11}\\
 & \equiv\sum_{n,m}\widetilde{v}_{lnm}e^{in\alpha_{1}}e^{im\alpha_{2}}\qquad(l=1,2)\nonumber 
\end{align}
The Fourier coefficient for spectral line $n\omega_{1}+m\omega_{2}$
is $\widetilde{v}_{lnm}=\sum_{j}a_{lj}\widetilde{w}_{jnm}$. We choose
$a_{lj}$ such that $\widetilde{v}_{110}=1$, $\widetilde{v}_{201}=1$,
and define $\widetilde{v}_{1nm}=\epsilon_{1nm}$, for all ${n,m}$
except ${n=1,m=0}$, and $\widetilde{v}_{2nm}=\epsilon_{2nm}$ for
all ${n,m}$ except for ${n=0,m=1}$. $\epsilon_{lnm}$ represents
fluctuation. Among all possible values for $a_{lj}$, the one with
minimized fluctuation most closely represents the rigid rotations.
In general, we have a minimization problem for a function $g_{0}$
quadratic in $a_{lj}$ with constraints $g_{1},g_{2}$:
\begin{equation}
\begin{aligned} & g_{0}(a_{lj})=\sum_{\tiny{\begin{matrix}n,m\\
|n-1|+|m|\neq0
\end{matrix}}}|\epsilon_{1nm}|^{2}+\sum_{\tiny{\begin{matrix}n,m\\
|n|+|m-1|\neq0
\end{matrix}}}|\epsilon_{2nm}|^{2}\\
 & g_{1}(a_{lj})=\widetilde{v}_{110}-1=0\\
 & g_{2}(a_{lj})=\widetilde{v}_{201}-1=0
\end{aligned}
\label{eq:12}
\end{equation}
If $g_{0}=0$, then $\epsilon_{lnm}$ are all zero, $v_{l}=e^{i\omega_{l}t}$
has a single frequency $\omega_{l}$, and $v_{1},v_{2}$ would be
exact pure rotations. In general the fluctuation would not vanish,
and for a finite power order $n_{s}$ of the square matrix (in this
paper we found $n_{s}=3$ would give very accurate solution) and eigenvector
number $n_{v}$ (in our example, we use $n_{v}=4$), we minimize the
fluctuation $g_{0}$ to improve the action-angle variables as follows. 

Use Lagrangian multiplier $\lambda_{1}$, $\lambda_{2}$, the minimization
problem is reduced to solving $2n_{v}+2$ linear equations for $2n_{v}+2$
unknown $a_{lj,}$$\lambda_{1},\lambda_{2}$:

{
\begin{equation}
\begin{aligned} & \frac{\partial g_{0}}{\partial a_{lj}}+\lambda_{1}\frac{\partial g_{1}}{\partial a_{lj}}+\lambda_{2}\frac{\partial g_{2}}{\partial a_{lj}}=0\qquad(l=1,2;\ j=1,2,..n_{v})\\
 & g_{1}=0,g_{2}=0
\end{aligned}
\label{eq:13}
\end{equation}
}The solution of Eq.(\ref{eq:13}) is straight forward, and gives
the linear combinations $a_{1k},a_{2k}$
\begin{align}
 & a_{1h}=\frac{\sum_{j}\left(F_{1}^{-1}\right)_{hj}\widetilde{w}_{j10}^{*}}{\sum_{m,j}\widetilde{w}_{m10}\left(F_{1}^{-1}\right)_{mj}\widetilde{w}_{j10}^{*}}\qquad(m,j,h=1,2,..n_{v})\nonumber \\
 & a_{2h}=\frac{\sum_{j}\left(F_{2}^{-1}\right)_{hj}\widetilde{w}_{j01}^{*}}{\sum_{m,j}\widetilde{w}_{m01}\left(F_{2}^{-1}\right)_{mj}\widetilde{w}_{j01}^{*}}\qquad\text{{with}}\label{eq:15}\\
 & \text{\ensuremath{\left(F_{1}\right)_{jh}\equiv\sum_{\tiny{\begin{matrix}n,m\\
 |n-1|+|m|\neq0 
\end{matrix}}}\widetilde{w}_{jnm}^{*}\widetilde{w}_{hnm}\qquad}(\ensuremath{m,n=0,1,2,}...,\ensuremath{n_{\theta}})}\nonumber \\
 & \text{\ensuremath{\left(F_{2}\right)_{jh}\equiv\sum_{\tiny{\begin{matrix}n,m\\
 |n|+|m-1|\neq0 
\end{matrix}}}\widetilde{w}_{jnm}^{*}\widetilde{w}_{hnm}\quad}}\nonumber 
\end{align}
For a given approximate trajectory $X^{(k-1)}(\alpha_{1},\alpha_{2})\equiv x^{(k-1)}$,$p_{x}^{(k-1)}$,$y^{(k-1)}$,$p_{y}^{(k-1)}$
as function of $\alpha_{1},\alpha_{2}$, the linear combinations $a_{1h}^{(k)},a_{2h}^{(k)}$
Eq.(\ref{eq:15}) determine the approximate action-angle variables
$v_{1}^{(k)},v_{2}^{(k)}$ with minimized fluctuation Eq.(\ref{eq:12}),
so they represent motion closer to pure rotations. Thus in every step
of the iteration, the new solution not only closer to an exact solution
of the exact dynamical equation Eq(\ref{eq:dynamics_theta}), it is also
closer to a pure rotation. With less fluctuation from pure rotation,
the convergence is faster. 

\bibliography{reference}

\begin{thebibliography}{32}%
\makeatletter
\providecommand \@ifxundefined [1]{%
 \@ifx{#1\undefined}
}%
\providecommand \@ifnum [1]{%
 \ifnum #1\expandafter \@firstoftwo
 \else \expandafter \@secondoftwo
 \fi
}%
\providecommand \@ifx [1]{%
 \ifx #1\expandafter \@firstoftwo
 \else \expandafter \@secondoftwo
 \fi
}%
\providecommand \natexlab [1]{#1}%
\providecommand \enquote  [1]{``#1''}%
\providecommand \bibnamefont  [1]{#1}%
\providecommand \bibfnamefont [1]{#1}%
\providecommand \citenamefont [1]{#1}%
\providecommand \href@noop [0]{\@secondoftwo}%
\providecommand \href [0]{\begingroup \@sanitize@url \@href}%
\providecommand \@href[1]{\@@startlink{#1}\@@href}%
\providecommand \@@href[1]{\endgroup#1\@@endlink}%
\providecommand \@sanitize@url [0]{\catcode `\\12\catcode `\$12\catcode
  `\&12\catcode `\#12\catcode `\^12\catcode `\_12\catcode `\%12\relax}%
\providecommand \@@startlink[1]{}%
\providecommand \@@endlink[0]{}%
\providecommand \url  [0]{\begingroup\@sanitize@url \@url }%
\providecommand \@url [1]{\endgroup\@href {#1}{\urlprefix }}%
\providecommand \urlprefix  [0]{URL }%
\providecommand \Eprint [0]{\href }%
\providecommand \doibase [0]{http://dx.doi.org/}%
\providecommand \selectlanguage [0]{\@gobble}%
\providecommand \bibinfo  [0]{\@secondoftwo}%
\providecommand \bibfield  [0]{\@secondoftwo}%
\providecommand \translation [1]{[#1]}%
\providecommand \BibitemOpen [0]{}%
\providecommand \bibitemStop [0]{}%
\providecommand \bibitemNoStop [0]{.\EOS\space}%
\providecommand \EOS [0]{\spacefactor3000\relax}%
\providecommand \BibitemShut  [1]{\csname bibitem#1\endcsname}%
\let\auto@bib@innerbib\@empty
\bibitem [{\citenamefont {Lichtenberg}\ and\ \citenamefont
  {Lieberman}(1992)}]{lieberman}%
  \BibitemOpen
  \bibfield  {author} {\bibinfo {author} {\bibfnamefont {A.~J.}\ \bibnamefont
  {Lichtenberg}}\ and\ \bibinfo {author} {\bibfnamefont {M.}~\bibnamefont
  {Lieberman}},\ }\href@noop {} {\emph {\bibinfo {title} {Regular and Chaotic
  Dynamics}}}\ (\bibinfo  {publisher} {Springer, New York},\ \bibinfo {year}
  {1992})\BibitemShut {NoStop}%
\bibitem [{\citenamefont {Ruth}(1987)}]{ruth}%
  \BibitemOpen
  \bibfield  {author} {\bibinfo {author} {\bibfnamefont {R.~D.}\ \bibnamefont
  {Ruth}},\ }\href {\doibase 10.1063/1.36365} {\bibfield  {journal} {\bibinfo
  {journal} {AIP Conf. Proc.}\ }\textbf {\bibinfo {volume} {153}},\ \bibinfo
  {pages} {150} (\bibinfo {year} {1987})}\BibitemShut {NoStop}%
\bibitem [{\citenamefont {Guignard}(1978)}]{guignard}%
  \BibitemOpen
  \bibfield  {author} {\bibinfo {author} {\bibfnamefont {G.}~\bibnamefont
  {Guignard}},\ }\href {\doibase 10.5170/CERN-1978-011} {\emph {\bibinfo
  {title} {A general treatment of resonances in accelerators}}},\ CERN Academic
  Training Lecture\ (\bibinfo  {publisher} {CERN},\ \bibinfo {address}
  {Geneva},\ \bibinfo {year} {1978})\ \bibinfo {note} {cERN, Geneva, 1977 -
  1978}\BibitemShut {NoStop}%
\bibitem [{\citenamefont {Schoch}(1958)}]{schoch}%
  \BibitemOpen
  \bibfield  {author} {\bibinfo {author} {\bibfnamefont {A.}~\bibnamefont
  {Schoch}},\ }\href {\doibase 10.5170/CERN-1957-021} {\emph {\bibinfo {title}
  {{Theory of linear and non-linear perturbations of betatron oscillations in
  alternating-gradient synchrotrons}}}},\ CERN Yellow Reports: Monographs\
  (\bibinfo  {publisher} {CERN},\ \bibinfo {address} {Geneva},\ \bibinfo {year}
  {1958})\BibitemShut {NoStop}%
\bibitem [{\citenamefont {Dragt}(1988)}]{dragt}%
  \BibitemOpen
  \bibfield  {author} {\bibinfo {author} {\bibfnamefont {A.~J.}\ \bibnamefont
  {Dragt}},\ }\href {\doibase 10.1063/1.37819} {\bibfield  {journal} {\bibinfo
  {journal} {AIP Conference Proceedings}\ }\textbf {\bibinfo {volume} {177}},\
  \bibinfo {pages} {261} (\bibinfo {year} {1988})},\ \Eprint
  {http://arxiv.org/abs/https://aip.scitation.org/doi/pdf/10.1063/1.37819}
  {https://aip.scitation.org/doi/pdf/10.1063/1.37819} \BibitemShut {NoStop}%
\bibitem [{\citenamefont {Berz}(1989)}]{berz}%
  \BibitemOpen
  \bibfield  {author} {\bibinfo {author} {\bibfnamefont {M.}~\bibnamefont
  {Berz}},\ }in\ \href {\doibase 10.1109/PAC.1989.73468} {\emph {\bibinfo
  {booktitle} {Proceedings of the 1989 IEEE Particle Accelerator Conference, .
  'Accelerator Science and Technology}}}\ (\bibinfo {year} {1989})\ pp.\
  \bibinfo {pages} {1419--1423 vol.3}\BibitemShut {NoStop}%
\bibitem [{\citenamefont {Chao}(2002)}]{chao}%
  \BibitemOpen
  \bibfield  {author} {\bibinfo {author} {\bibfnamefont {A.}~\bibnamefont
  {Chao}},\ }\href@noop {} {\enquote {\bibinfo {title} {{Lecture notes on
  topics in accelerator physics}},}\ } (\bibinfo {year} {2002})\BibitemShut
  {NoStop}%
\bibitem [{\citenamefont {Bazzani}\ \emph {et~al.}(1994)\citenamefont
  {Bazzani}, \citenamefont {Servizi}, \citenamefont {Todesco},\ and\
  \citenamefont {Turchetti}}]{bazzani}%
  \BibitemOpen
  \bibfield  {author} {\bibinfo {author} {\bibfnamefont {A.}~\bibnamefont
  {Bazzani}}, \bibinfo {author} {\bibfnamefont {G.}~\bibnamefont {Servizi}},
  \bibinfo {author} {\bibfnamefont {E.}~\bibnamefont {Todesco}}, \ and\
  \bibinfo {author} {\bibfnamefont {G.}~\bibnamefont {Turchetti}},\ }\href
  {\doibase 10.5170/CERN-1994-002} {\emph {\bibinfo {title} {{A normal form
  approach to the theory of nonlinear betatronic motion}}}},\ CERN Yellow
  Reports: Monographs\ (\bibinfo  {publisher} {CERN},\ \bibinfo {address}
  {Geneva},\ \bibinfo {year} {1994})\BibitemShut {NoStop}%
\bibitem [{\citenamefont {Forest}(1998)}]{forest}%
  \BibitemOpen
  \bibfield  {author} {\bibinfo {author} {\bibfnamefont {E.}~\bibnamefont
  {Forest}},\ }\href@noop {} {\emph {\bibinfo {title} {Beam Dynamics: A New
  Attitude and Framework}}}\ (\bibinfo  {publisher} {Harwood, Amsterdam,
  Netherlands},\ \bibinfo {year} {1998})\BibitemShut {NoStop}%
\bibitem [{\citenamefont {Forest}\ \emph {et~al.}(1989)\citenamefont {Forest},
  \citenamefont {Berz},\ and\ \citenamefont {Irwin}}]{forest1}%
  \BibitemOpen
  \bibfield  {author} {\bibinfo {author} {\bibfnamefont {E.}~\bibnamefont
  {Forest}}, \bibinfo {author} {\bibfnamefont {M.}~\bibnamefont {Berz}}, \ and\
  \bibinfo {author} {\bibfnamefont {J.}~\bibnamefont {Irwin}},\ }\href
  {https://cds.cern.ch/record/1053511} {\bibfield  {journal} {\bibinfo
  {journal} {Part. Accel.}\ }\textbf {\bibinfo {volume} {24}},\ \bibinfo
  {pages} {91} (\bibinfo {year} {1989})}\BibitemShut {NoStop}%
\bibitem [{\citenamefont {Michelotti}\ and\ \citenamefont
  {Lifshitz}(1995)}]{michelotti}%
  \BibitemOpen
  \bibfield  {author} {\bibinfo {author} {\bibfnamefont {L.}~\bibnamefont
  {Michelotti}}\ and\ \bibinfo {author} {\bibfnamefont {E.~M.}\ \bibnamefont
  {Lifshitz}},\ }\href@noop {} {\emph {\bibinfo {title} {IntermediateClassical
  Dynamics with Applications to Beam Physics}}}\ (\bibinfo  {publisher} {Wiley,
  New York},\ \bibinfo {year} {1995})\BibitemShut {NoStop}%
\bibitem [{\citenamefont {Yu}\ and\ \citenamefont {Nash}()}]{yu1}%
  \BibitemOpen
  \bibfield  {author} {\bibinfo {author} {\bibfnamefont {L.-H.}\ \bibnamefont
  {Yu}}\ and\ \bibinfo {author} {\bibfnamefont {B.}~\bibnamefont {Nash}},\ }in\
  \href {https://jacow.org/PAC2009/papers/TH6PFP067.pdf} {\emph {\bibinfo
  {booktitle} {Proc. PAC'09}}}\ (\bibinfo  {publisher} {JACoW Publishing,
  Geneva, Switzerland})\ pp.\ \bibinfo {pages} {3862--3864}\BibitemShut
  {NoStop}%
\bibitem [{\citenamefont {Yu}(2017)}]{yu2}%
  \BibitemOpen
  \bibfield  {author} {\bibinfo {author} {\bibfnamefont {L.~H.}\ \bibnamefont
  {Yu}},\ }\href {\doibase 10.1103/PhysRevAccelBeams.20.034001} {\bibfield
  {journal} {\bibinfo  {journal} {Phys. Rev. Accel. Beams}\ }\textbf {\bibinfo
  {volume} {20}},\ \bibinfo {pages} {034001} (\bibinfo {year}
  {2017})}\BibitemShut {NoStop}%
\bibitem [{\citenamefont {Yu}\ \emph {et~al.}()\citenamefont {Yu},
  \citenamefont {Hao}, \citenamefont {Hidaka}, \citenamefont {Plassard},\ and\
  \citenamefont {Smaluk}}]{yu3}%
  \BibitemOpen
  \bibfield  {author} {\bibinfo {author} {\bibfnamefont {L.~H.}\ \bibnamefont
  {Yu}}, \bibinfo {author} {\bibfnamefont {Y.}~\bibnamefont {Hao}}, \bibinfo
  {author} {\bibfnamefont {Y.}~\bibnamefont {Hidaka}}, \bibinfo {author}
  {\bibfnamefont {F.}~\bibnamefont {Plassard}}, \ and\ \bibinfo {author}
  {\bibfnamefont {V.~V.}\ \bibnamefont {Smaluk}},\ }in\ \href {\doibase
  10.18429/JACoW-IPAC2021-MOPAB041} {\emph {\bibinfo {booktitle} {Proc.
  IPAC'21}}}\ (\bibinfo  {publisher} {JACoW Publishing, Geneva, Switzerland})\
  pp.\ \bibinfo {pages} {182--185}\BibitemShut {NoStop}%
\bibitem [{\citenamefont {Hao}\ \emph {et~al.}()\citenamefont {Hao},
  \citenamefont {Anderson},\ and\ \citenamefont {Yu}}]{hao}%
  \BibitemOpen
  \bibfield  {author} {\bibinfo {author} {\bibfnamefont {Y.}~\bibnamefont
  {Hao}}, \bibinfo {author} {\bibfnamefont {K.~J.}\ \bibnamefont {Anderson}}, \
  and\ \bibinfo {author} {\bibfnamefont {L.~H.}\ \bibnamefont {Yu}},\ }in\
  \href {\doibase 10.18429/JACoW-IPAC2021-THPAB016} {\emph {\bibinfo
  {booktitle} {Proc. IPAC'21}}}\ (\bibinfo  {publisher} {JACoW Publishing,
  Geneva, Switzerland})\ pp.\ \bibinfo {pages} {3788--3791}\BibitemShut
  {NoStop}%
\bibitem [{\citenamefont {Dierker}(2007)}]{NSLSII}%
  \BibitemOpen
  \bibfield  {author} {\bibinfo {author} {\bibfnamefont {S.}~\bibnamefont
  {Dierker}},\ }\href {\doibase 10.2172/1010602} {\enquote {\bibinfo {title}
  {Nsls-ii preliminary design report},}\ } (\bibinfo {year} {2007})\BibitemShut
  {NoStop}%
\bibitem [{\citenamefont {Nadolski}\ and\ \citenamefont
  {Laskar}(2003)}]{laskar}%
  \BibitemOpen
  \bibfield  {author} {\bibinfo {author} {\bibfnamefont {L.}~\bibnamefont
  {Nadolski}}\ and\ \bibinfo {author} {\bibfnamefont {J.}~\bibnamefont
  {Laskar}},\ }\href {\doibase 10.1103/PhysRevSTAB.6.114801} {\bibfield
  {journal} {\bibinfo  {journal} {Phys. Rev. ST Accel. Beams}\ }\textbf
  {\bibinfo {volume} {6}},\ \bibinfo {pages} {114801} (\bibinfo {year}
  {2003})}\BibitemShut {NoStop}%
\bibitem [{PyT()}]{PyTPSA}%
  \BibitemOpen
  \href@noop {} {}\bibinfo {note}
  {\url{https://github.com/YueHao/PyTPSA}}\BibitemShut {NoStop}%
\bibitem [{\citenamefont {Wayne}(1990)}]{wayne}%
  \BibitemOpen
  \bibfield  {author} {\bibinfo {author} {\bibfnamefont {C.~E.}\ \bibnamefont
  {Wayne}},\ }\href {\doibase cmp/1104180217} {\bibfield  {journal} {\bibinfo
  {journal} {Communications in Mathematical Physics}\ }\textbf {\bibinfo
  {volume} {127}},\ \bibinfo {pages} {479 } (\bibinfo {year}
  {1990})}\BibitemShut {NoStop}%
\bibitem [{\citenamefont {K\r{a}gstr\"{o}m}\ and\ \citenamefont
  {Ruhe}(1980)}]{Kagstrom1}%
  \BibitemOpen
  \bibfield  {author} {\bibinfo {author} {\bibfnamefont {B.}~\bibnamefont
  {K\r{a}gstr\"{o}m}}\ and\ \bibinfo {author} {\bibfnamefont {A.}~\bibnamefont
  {Ruhe}},\ }\href {\doibase 10.1145/355900.355917} {\bibfield  {journal}
  {\bibinfo  {journal} {ACM Trans. Math. Softw.}\ }\textbf {\bibinfo {volume}
  {6}},\ \bibinfo {pages} {437–443} (\bibinfo {year} {1980})}\BibitemShut
  {NoStop}%
\bibitem [{\citenamefont {K{\aa}gstr{\"o}m}\ and\ \citenamefont
  {Ruhe}(1980)}]{Kagstrom2}%
  \BibitemOpen
  \bibfield  {author} {\bibinfo {author} {\bibfnamefont {B.}~\bibnamefont
  {K{\aa}gstr{\"o}m}}\ and\ \bibinfo {author} {\bibfnamefont {A.}~\bibnamefont
  {Ruhe}},\ }\href@noop {} {\bibfield  {journal} {\bibinfo  {journal} {ACM
  Trans. Math. Softw.}\ }\textbf {\bibinfo {volume} {6}},\ \bibinfo {pages}
  {398} (\bibinfo {year} {1980})}\BibitemShut {NoStop}%
\bibitem [{\citenamefont {Broer}(2004)}]{broer}%
  \BibitemOpen
  \bibfield  {author} {\bibinfo {author} {\bibfnamefont {H.}~\bibnamefont
  {Broer}},\ }\href {\doibase 10.1090/S0273-0979-04-01009-2} {\bibfield
  {journal} {\bibinfo  {journal} {Bulletin of the American Mathematical
  Society}\ }\textbf {\bibinfo {volume} {41}},\ \bibinfo {pages} {1} (\bibinfo
  {year} {2004})}\BibitemShut {NoStop}%
\bibitem [{\citenamefont {Arnold}(2009)}]{arnold}%
  \BibitemOpen
  \bibfield  {author} {\bibinfo {author} {\bibfnamefont {V.}~\bibnamefont
  {Arnold}},\ }\enquote {\bibinfo {title} {Small denominators. i. mapping of
  the circumference onto itself},}\ \ (\bibinfo {year} {2009})\BibitemShut
  {NoStop}%
\bibitem [{\citenamefont {Borland}(2000)}]{borland}%
  \BibitemOpen
  \bibfield  {author} {\bibinfo {author} {\bibfnamefont {M.}~\bibnamefont
  {Borland}},\ }in\ \href {\doibase 10.2172/761286} {\emph {\bibinfo
  {booktitle} {{6th International Computational Accelerator Physics Conference
  (ICAP 2000)}}}}\ (\bibinfo {year} {2000})\BibitemShut {NoStop}%
\bibitem [{\citenamefont {Deb}(2001)}]{MOGA}%
  \BibitemOpen
  \bibfield  {author} {\bibinfo {author} {\bibfnamefont {K.}~\bibnamefont
  {Deb}},\ }\enquote {\bibinfo {title} {Multiobjective optimization using
  evolutionary algorithms. wiley, new york},}\ \ (\bibinfo {year}
  {2001})\BibitemShut {NoStop}%
\bibitem [{\citenamefont {Yang}\ \emph {et~al.}(2011)\citenamefont {Yang},
  \citenamefont {Li}, \citenamefont {Guo},\ and\ \citenamefont
  {Krinsky}}]{yang}%
  \BibitemOpen
  \bibfield  {author} {\bibinfo {author} {\bibfnamefont {L.}~\bibnamefont
  {Yang}}, \bibinfo {author} {\bibfnamefont {Y.}~\bibnamefont {Li}}, \bibinfo
  {author} {\bibfnamefont {W.}~\bibnamefont {Guo}}, \ and\ \bibinfo {author}
  {\bibfnamefont {S.}~\bibnamefont {Krinsky}},\ }\href {\doibase
  10.1103/PhysRevSTAB.14.054001} {\bibfield  {journal} {\bibinfo  {journal}
  {Phys. Rev. ST Accel. Beams}\ }\textbf {\bibinfo {volume} {14}},\ \bibinfo
  {pages} {054001} (\bibinfo {year} {2011})}\BibitemShut {NoStop}%
\bibitem [{\citenamefont {Fortin}\ \emph {et~al.}(2012)\citenamefont {Fortin},
  \citenamefont {De~Rainville}, \citenamefont {Gardner}, \citenamefont
  {Parizeau},\ and\ \citenamefont {Gagn\'{e}}}]{fortin}%
  \BibitemOpen
  \bibfield  {author} {\bibinfo {author} {\bibfnamefont {F.-A.}\ \bibnamefont
  {Fortin}}, \bibinfo {author} {\bibfnamefont {F.-M.}\ \bibnamefont
  {De~Rainville}}, \bibinfo {author} {\bibfnamefont {M.-A.~G.}\ \bibnamefont
  {Gardner}}, \bibinfo {author} {\bibfnamefont {M.}~\bibnamefont {Parizeau}}, \
  and\ \bibinfo {author} {\bibfnamefont {C.}~\bibnamefont {Gagn\'{e}}},\
  }\href@noop {} {\bibfield  {journal} {\bibinfo  {journal} {J. Mach. Learn.
  Res.}\ }\textbf {\bibinfo {volume} {13}},\ \bibinfo {pages} {2171–2175}
  (\bibinfo {year} {2012})}\BibitemShut {NoStop}%
\bibitem [{DEA()}]{DEAP2}%
  \BibitemOpen
  \href@noop {} {}\bibinfo {note}
  {\url{https://github.com/DEAP/deap}}\BibitemShut {NoStop}%
\bibitem [{\citenamefont {J.~Bengtsson}()}]{tracy}%
  \BibitemOpen
  \bibfield  {author} {\bibinfo {author} {\bibfnamefont {H.~N.}\ \bibnamefont
  {J.~Bengtsson}, \bibfnamefont {E.Forest}},\ }\href@noop {} {}\bibinfo {note}
  {Tracy User Manual}\BibitemShut {NoStop}%
\bibitem [{\citenamefont {Skowronski}\ \emph {et~al.}()\citenamefont
  {Skowronski}, \citenamefont {Forest}, \citenamefont {Schmidt},\ and\
  \citenamefont {de~Maria}}]{skowronski}%
  \BibitemOpen
  \bibfield  {author} {\bibinfo {author} {\bibfnamefont {P.~K.}\ \bibnamefont
  {Skowronski}}, \bibinfo {author} {\bibfnamefont {E.}~\bibnamefont {Forest}},
  \bibinfo {author} {\bibfnamefont {F.}~\bibnamefont {Schmidt}}, \ and\
  \bibinfo {author} {\bibfnamefont {R.}~\bibnamefont {de~Maria}},\ }in\ \href
  {https://jacow.org/icap06/papers/WEPPP12.pdf} {\emph {\bibinfo {booktitle}
  {Proc. ICAP'06}}}\ (\bibinfo  {publisher} {JACoW Publishing, Geneva,
  Switzerland})\ pp.\ \bibinfo {pages} {209--212}\BibitemShut {NoStop}%
\bibitem [{sqm()}]{sqmxcode}%
  \BibitemOpen
  \href@noop {} {}\bibinfo {note}
  {\url{https://github.com/yhidaka/squarematrix}}\BibitemShut {NoStop}%
\bibitem [{\citenamefont {Brown}\ and\ \citenamefont {Neumann}(1977)}]{brown}%
  \BibitemOpen
  \bibfield  {author} {\bibinfo {author} {\bibfnamefont {M.}~\bibnamefont
  {Brown}}\ and\ \bibinfo {author} {\bibfnamefont {W.~D.}\ \bibnamefont
  {Neumann}},\ }\href {\doibase 10.1307/mmj/1029001816} {\bibfield  {journal}
  {\bibinfo  {journal} {Michigan Mathematical Journal}\ }\textbf {\bibinfo
  {volume} {24}},\ \bibinfo {pages} {21 } (\bibinfo {year} {1977})}\BibitemShut
  {NoStop}%
\end{thebibliography}%


%

\end{document}